\documentclass[longauth]{aa} % for the long lists of affiliations 
\usepackage{graphicx}
\usepackage{lipsum} 
\usepackage{txfonts}
\usepackage{amssymb}
\usepackage{amsmath}
\usepackage{sidecap}
\usepackage{multirow}
\usepackage{xcolor}
\usepackage{natbib}
\bibpunct{(}{)}{;}{a}{}{,}
\begin{document}

   \title{HD 117214 debris disk: scattered-light images and constraints on the presence of planets
   \thanks{Based on data collected at the European Southern Observatory, Chile under program 1100.C-0481.} }
   \author{N.~Engler\inst{\ref{instch1}} 
   \and C.~Lazzoni\inst{\ref{insti1}}  
   \and R.~Gratton\inst{\ref{insti1}}
   \and J.~Milli\inst{\ref{insteso2}}  
   \and H.M.~Schmid\inst{\ref{instch1}} 
   \and G.~Chauvin\inst{\ref{instf1},\ref{instcl1}}    
   \and Q.~Kral\inst{\ref{instf4}}     
   \and N.~Pawellek\inst{\ref{insten1},\ref{instun1}} 
   \and P.~Th\'{e}bault\inst{\ref{instf4}}
   \and A.~Boccaletti\inst{\ref{instf4}}
   \and M.~Bonnefoy\inst{\ref{instf1}} 
   \and S.~Brown\inst{\ref{instf1}} 
   \and T. Buey\inst{\ref{instf4}}
   \and F.~Cantalloube\inst{\ref{instf1}}
   \and M.~Carle\inst{\ref{instf3}}
   \and A.~Cheetham\inst{\ref{instd1}}
   \and S.~Desidera\inst{\ref{insti1}}
   \and M.~Feldt\inst{\ref{instd1}}   
   \and C.~Ginski\inst{\ref{instnl2}} 
   \and D.~Gisler\inst{\ref{instd2}}  
   \and Th.~Henning\inst{\ref{instd1}}  
   \and S.~Hunziker\inst{\ref{instch1}}    
   \and A.M.~Lagrange\inst{\ref{instf1}}    
   \and M.~Langlois\inst{\ref{instf3},\ref{instf7}}    
   \and D.~Mesa\inst{\ref{insti1}} 
   \and M.R.~Meyer\inst{\ref{instch1},\ref{instam1}}
   \and O.~Moeller-Nilsson\inst{\ref{instd1}}
   \and J.~Olofsson\inst{\ref{instcl1},\ref{instd1},\ref{instcl2}} 
   \and C.~Petit\inst{\ref{instf6}}
   \and S.~Petrus\inst{\ref{instf3}}
   \and S.P.~Quanz\inst{\ref{instch1}}  
   \and E.~Rickman\inst{\ref{instch2}}
   \and E.~Stadler\inst{\ref{instf1}}
   \and T.~Stolker\inst{\ref{instch1}}
   \and A.~Vigan\inst{\ref{instf3}}
   \and F.~Wildi\inst{\ref{instch2}}  
   \and A.~Zurlo\inst{\ref{instcl3},\ref{instcl4},\ref{instf3}}  
             }

\institute{
ETH Zurich, Institute for Particle Physics and Astrophysics, 
Wolfgang-Pauli-Strasse 27, 
CH-8093 Zurich, Switzerland \\ \email{englern@phys.ethz.ch}\label{instch1}
\and
INAF – Osservatorio Astronomico di Padova, Vicolo
dell’Osservatorio 5, 35122 Padova, Italy\label{insti1}
\and
CNRS, IPAG, Universit\'{e} Grenoble Alpes, IPAG, 38000 Grenoble, France\label{instf1}
\and
LESIA, Observatoire de Paris, Universit\'{e} PSL, CNRS, Sorbonne
Universit{\'e}, Universit{\'e} de Paris, 5 place Jules Janssen, 92195 Meudon, France\label{instf4}
\and
European Southern Observatory, Alonso de Cordova 3107, Casilla
19001 Vitacura, Santiago 19, Chile\label{insteso2}
\and
Institute of Astronomy, University of Cambridge, Madingley Road, Cambridge CB3 0HA, UK\label{insten1}
\and
IKonkoly Observatory, Research Centre for Astronomy and Earth Sciences, Konkoly-Thege Miklós út 15-17, H-1121 Budapest, Hungary\label{instun1}
\and
Max-Planck-Institut f\"{u}r Astronomie, K\"{o}nigstuhl 17, 69117
Heidelberg, Germany\label{instd1}
\and
Aix Marseille Universit\'e, CNRS, LAM - Laboratoire d'Astrophysique de Marseille, UMR 7326, 13388, Marseille, France\label{instf3}
\and
Anton Pannekoek Astronomical Institute, University of Amsterdam,
PO Box 94249, 1090 GE Amsterdam, The Netherlands\label{instnl2}
\and
Kiepenheuer-Institut f\"{u}r Sonnenphysik, Schneckstr. 6, D-79104
Freiburg, Germany\label{instd2}
\and
Geneva Observatory, University of Geneva, Chemin des Mailettes
51, 1290 Versoix, Switzerland\label{instch2}
\and
Centre de Recherche Astrophysique de Lyon, CNRS/ENSL
Universit\'{e} Lyon 1, 9 av. Ch. Andr\'{e}, 69561 Saint-Genis-Laval,
France\label{instf7}
\and
Department of Astronomy, University of Michigan, 311 West Hall, 1085 S. University Avenue, Ann Arbor, MI 48109, USA\label{instam1}
\and
Instituto de F\'isica y Astronom\'ia, Facultad de Ciencias, Universidad de Valpara\'iso, Av. Gran Breta\~na 1111, Playa Ancha, Valpara\'iso, Chile\label{instcl1}
\and
N\'ucleo Milenio Formaci\'on Planetaria - NPF, Universidad de Valpara\'iso, Av. Gran Breta\~na 1111, Valpara\'iso, Chile\label{instcl2}
\and
$^{}$N\'ucleo de Astronom\'ia, Facultad de Ingenier\'ia y Ciencias, Universidad Diego Portales, Av. Ejercito 441, Santiago, Chile\label{instcl3}
\and
Escuela de Ingenier\'ia Industrial, Facultad de Ingenier\'ia y Ciencias, Universidad Diego Portales, Av. Ejercito 441, Santiago, Chile\label{instcl4}
\and
ONERA, The French Aerospace Lab BP72, 29 avenue de la
Division Leclerc, 92322 Ch\^{a}tillon Cedex, France\label{instf6}
            }
            
\abstract
%context
{Young stars with debris disks are the most promising targets for an exoplanet search because debris indicate a successful formation of planetary bodies.  
Debris disks can be shaped by planets into ring structures that give valuable indications on
the presence and location of planets in the disk.} 
% aim
{We performed observations of the Sco-Cen F star HD 117214 to search for planetary companions and to characterize the debris disk structure. }
% method
{HD 117214 was observed with the SPHERE subsystems IRDIS, IFS, and ZIMPOL at optical and near-IR wavelengths using angular and polarimetric differential imaging techniques. This provided the first images of scattered light from the debris disk with the highest spatial resolution of 25 mas and the inner working angle $< 0.1''$. With the observations with IRDIS and IFS we derived detection limits for substellar companions. The geometrical parameters of the detected disk were constrained by fitting 3D models for the
scattering of an optically thin dust disk. Investigating the possible origin of the disk gap, we introduced putative planets therein and modeled the planet-disk and planet-planet dynamical interactions. The obtained planetary architectures were compared with the detection limit curves. }
% results
{The debris disk has an axisymmetric ring structure with a radius of $0.42(\pm 0.01)''$ or $\sim$45 au and an inclination of $71(\pm 2.5)^\circ$ and exhibits a $0.4''$ ($\sim$40 au) wide inner cavity. From the polarimetric data, we derive a polarized flux contrast for the disk of $(F_{\rm pol})_{\rm disk}/F_{\rm \ast}> (3.1 \pm 1.2)\cdot 10^{-4}$ in the RI band.}  
% Conclusions
{The fractional scattered polarized flux of the disk is eight times lower than the fractional IR flux excess. This ratio is similar to the one obtained for the debris disk HIP 79977, indicating that dust radiation properties are similar for these two disks. \\
Inside the disk cavity we achieve high-sensitivity limits on planetary companions with a mass down to $\sim$ 4~$M_{\rm J}$ at projected radial separations between 0.2$''$ and 0.4$''$. We can exclude stellar companions at a radial separation larger than 75 mas from the star.
}
%  \abstract
  % context heading (optional)
  % {} leave it empty if necessary  


\authorrunning{Engler et al.}

\titlerunning{HD 117214 debris disk}

   \maketitle
%
%________________________________________________________________

\section{Introduction}
Circumstellar debris disks around young stars ($\sim$1$0-100$ Myr) are often considered to be the remains of protoplanetary disks and are seen as a direct evidence for the presence of large planetesimals and planets because the large amount of dust observed in these stellar systems is thought to be generated in destructive collisions between large solid bodies \citep[e.g.,][and references therein]{Wyatt2008, Krivov2010, Hughes2018}. When they orbit a star, planets scatter planetesimals away and gravitationally attract small rocks and tiny dust grains. This clears out large surrounding areas around the planets \citep[e.g.,][and references therein]{Faber2007, Dipierro2016, Geiler2017}. 
In this way, planets can create wide empty gaps in dusty disks and shape them into ring structures. This scenario for the evolution of a planetary system provides one possible explanation for the multiple concentric rings observed in protoplanetary disks \citep[e.g.,][]{Andrews2018} and debris disks \citep{Golimowski2011, Perrot2016, Feldt2017, Bonnefoy2017, Marino2018, Engler2019, Boccaletti2019}. Large amount of gas in the disks \citep{Kral2019} can also lead to similar results \citep{Lyra2013, Richert2018}. However, the idea of a planetary origin of the ring structure is supported by the planets that have been discovered in the gaps of protoplanetary disks \citep{Keppler2018, Haffert2019} and between two debris belts \citep{Marois2010, Rameau2013}. For this reason, debris disks, especially those that are supposed to consist of at least two planetesimal belts \citep{Lazzoni2018, Kennedy2014}, are the primary targets in searches for extrasolar planets. 

The Scorpius-Centaurus OB association (Sco-Cen) is one preferred region for surveys searching for debris disks and young exoplanets near the Sun. The region is divided into three large subgroups and contains hundreds of young stars located at distances of $\sim 100-200$ pc \citep{GaiaCollaboration2018}. The F6V star \citep{Houk1975} HD 117214 (HIP 65875) we discuss here is a member of the Lower Centaurus Crux subgroup with an estimated age of $\sim$17 Myr \citep{Mamajek2002}.

\begin{table*} 
      \caption[]{Log of IRDIS / IFS observations with atmospheric conditions.}
      \centering
         \label{t_Settings_IRDIS}
                \renewcommand{\arraystretch}{1.3}
         \begin{tabular}{ccccccccccc}
            \hline 
            \hline
            \multirow{ 3}{*}{Date} & \multirow{ 3}{2cm}{Observation identification$^1$}& Field & Total exposure &  & \multicolumn{4}{c}{Observing conditions$^2$} \\\cline{6-9} 
            & & rotation & time &  & Airmass & Seeing & Coherence time & Wind speed \\
             & & [$^\circ$]& [min] &  &  & [$''$] & [ms] & [ms$^{-1}$] \\
            \hline
            \hline
            \noalign{\smallskip}
      2019-03-11&OBS070\_0084-0099 & 31 & 76.8  & &1.23--1.21&  $0.41 \pm 0.05$ & $8.1 \pm 1.6$ & $5 \pm 0$\\
            \hline
            \hline
            \noalign{\smallskip}
            
\end{tabular}
\tablefoot{
\tablefoottext{1}{The  observation  identification  corresponds  to  the  fits-file  header  keyword ``origname''  without the prefix ``SPHERE\_IRDIFS\_IRDIS\_'' or ``SPHERE\_IRDIFS\_IFS\_''. The first three digits give the day of the year followed by the four-digit observation number.}
\tablefoottext{2}{For seeing condition, coherence time, and wind speed, the mean with standard deviation of the distribution are given.}
}
 \end{table*}   

\begin{table*} 
      \caption[]{Log of ZIMPOL observations with the atmospheric conditions for each run.}
      \centering
         \label{t_Settings}
                \renewcommand{\arraystretch}{1.3}
         \begin{tabular}{ccccccccccc}
            \hline 
            \hline
 %           \noalign{\smallskip}
            \multirow{ 3}{*}{Date} & \multirow{ 3}{2cm}{Observation identification$^1$}& Field & Total exposure &  & \multicolumn{4}{c}{Observing conditions$^2$} \\\cline{6-9} 
            & & offset & time in FP / SP &  & Airmass & Seeing & Coherence time & Wind speed \\
             & & [$^\circ$]& [min] &  &  & [$''$] & [ms] & [ms$^{-1}$] \\
            \hline
            \hline
            \noalign{\smallskip}
      2018-02-28&OBS059\_0002-0049 & 0 & 4.4 / 32  & &1.29--1.23&  $0.93 \pm 0.17$ & $3.0 \pm 0.5$ & $10 \pm 1$\\
      2018-02-28&OBS059\_0050-0097 & 60 & 4.4 / 32 & &1.23--1.21&  $1.38 \pm 0.30$ & $2.6 \pm 0.4$ & $10 \pm 1$\\
      2018-06-22&OBS173\_0001-0028 & 60 & 2.3 / 16 & &1.21--1.21&  $0.93 \pm 0.05$ & $4.1 \pm 0.2$ & $9 \pm 0$\\
            \hline
            \hline
            \noalign{\smallskip}
            
\end{tabular}

\tablefoot{
\tablefoottext{1}{The  observation  identification  corresponds  to  the  fits-file  header  keyword ``origname''  without the prefix ``SPHERE\_ZIMPOL\_''. }
\tablefoottext{2}{For seeing condition, coherence time, and wind speed, the mean with standard deviation of the distribution are given. }
}
\end{table*}   

The star is located at a distance of $107.6 \pm 0.5$ pc \citep{GaiaCollaboration2018}. A high-IR excess indicating circumstellar dust around HD 117214 was detected with the \textit{Spitzer} telescope \citep{Chen2011}. \cite{Lieman_Sifry_2016} have observed this debris disk with ALMA at 1.24 mm and measured a flux of $270 \pm 50$ mJy, but with their spatial resolution of $1.32 \times 0.86$ arcsec the disk was not resolved. They also searched for CO emission that was found to be lower than their 3$\sigma$ upper limit of 39 mJy~km~s$^{-1}$. 

In 2018 and 2019, HD 117214 was observed with the Spectro-Polarimetric High-contrast Exoplanet REsearch (SPHERE) instrument \citep{Beuzit2019} at the Very Large Telescope (VLT) in Chile in the course of the guaranteed-time observation (GTO) programs SpHere INfrared survey for Exoplanets (SHINE) and SPHERE-DISK. This work presents these observations and describes the first scattered-light images of the debris disk around HD 117214
at different wavelengths (visual to near-IR) and the detection limits for stellar and substellar companions.
The following Sects.~\ref{s_Observations} and \ref{s_data} discuss the observations and data reduction. In Sect.~\ref{s_Results} we analyze the morphology of the disk observed in the total and polarized intensity data and present the results of modeling the disk geometry. Section~\ref{s_Photometry} is dedicated to the photometric analysis of the data. In Sect.~\ref{Discussion} we compare the disk HD 117214 with another Sco-Cen debris disk, HIP 79977 \citep{Engler2017}, discuss the similarity between the scattering phase function (SPF) that we obtained for the HD 117214 disk and SPFs measured for other debris disks, and investigate the possible presence of giant planets inside the debris belt. We conclude and summarize our results in Sect.~\ref{s_Summary}.

%__________________________________________________________________

\section{Observations} \label{s_Observations}
\subsection{IRDIS / IFS observations}
HD 117214 was observed on 2019 March 11 simultaneously with the Infra-Red Dual-beam Imager and Spectrograph \citep[IRDIS,][]{Dohlen2008} and the Integral Field Spectrograph \citep[IFS, ][]{Claudi2008}. The observations were performed in the IRDIFS-EXT pupil-stabilized mode \citep{Zurlo2014} using IRDIS in the dual-band imaging mode \citep[DBI,][]{Vigan2010} with the K1K2 filters ($\lambda_{\rm K1}=2.110\,\mu$m, $\Delta\lambda_{\rm K1} = 0.102\,\mu$m; $\lambda_{\rm K2}=2.251\,\mu$m, $\Delta\lambda_{\rm K2} = 0.109\,\mu$m) and the IFS in \textit{Y-H} mode ($0.97-1.66\, \mu$m, $R_\lambda=35$). The field of view (FOV) of the IRDIS detector is approximately $11'' \times 12.5''$, and that of the IFS is $1.73'' \times 1.73''$. An apodized Lyot coronagraph N\_ALC\_Ks \citep[diameter of 240~mas,][]{Carbillet2011, Guerri2011} was used to block the stellar light. The recorded sequence consists of $48 \times 96$ s individual exposures, yielding a total integration time of 76.8 min and covering a field rotation of 31$^\circ$ for both instruments. 

To measure the stellar flux, several short exposures, where the star was offset from the coronagraphic
mask, were taken before and after the science sequence using a neutral density filter ND1.0 with a transmission of about 85\%. The detector integration time (DIT) of these flux calibration frames is $\approx$~2~s (DIT = 4~s for the IFS). 

Additionally, a ``center frame'' was taken at the beginning of the science observation using the deformable mirror waffle mode \citep{Langlois2013}. This frame provides a measurement of the star position behind the coronagraph with an accuracy of up to 0.1 pixel or 1.2 mas \citep{Vigan2016}. The stability of the stellar position during the observation is ensured by the differential tip-tilt sensor \citep[DTTS,][]{Baudoz2010}. The observations were performed under excellent observing conditions with average seeing of $0.41''$ and coherence time of $8.1$~ms (see Table~\ref{t_Settings_IRDIS}).

\subsection{ZIMPOL observations}
The polarimetric observations of HD 117214 with SPHERE-ZIMPOL \citep[Zurich IMaging POLarimeter;][]{Schmid2018} were carried out on 2018 February 28 and June 22.
Images were taken in the RI band (hereafter Very Broad Band, or VBB), which covers the wavelength range of the R and I bands ($\lambda_c=735$~nm, $\Delta\lambda=290$~nm). This filter provides the highest throughput of photons, which is useful for the detection of a faint target such as a debris disk.

The measurements were performed in the polarimetric field-stabilized mode P2 of ZIMPOL using the fast-polarimetry (FP) and slow-polarimetry (SP) detector modes. The terms ``fast'' and ``slow'' refer to the cycle frequency for polarimetric modulation and demodulation. In ZIMPOL, the signal is modulated with a ferro-electric
liquid crystal retarder and a polarization beam splitter and sent to two demodulating CCD detectors/cameras (cam 1 and cam 2). Intensities of two opposite polarization states, one perpendicular to the ZIMPOL bench ($I_\perp$) and the other one parallel to it ($I_\parallel$), are measured quasi-simultaneously by each detector. The FP mode with a high cycle frequency of 967.5~Hz and high detector gain of 10.5 e$^-$/ADU allows for the measurements of bright sources with short integration times without detector saturation. The SP mode has a cycle frequency of 26.97~Hz and a lower detector gain of~1.5 e$^-$/ADU, and a much lower read-out noise level than the FP mode. This provides a higher sensitivity when many exposures with $t_{\rm DIT} \geqq 10$ s are added at radial separations larger than approximately 0.12$''$. Such long exposures saturate the detector for the peak of the point spread function (PSF), but the saturation can be avoided when the coronagraph is used. 

The successful detection of a faint polarimetric signal with ZIMPOL requires a long integration time, a very accurate image centering, and a correction for the differential polarimetric beam shift introduced by inclined mirrors \citep{Schmid2018}. This beam shift varies with the sky position and instrument configuration (filter, derotator mode), and a correction requires determining the stellar PSF peak position in both images $I_\perp$ and $I_\parallel$ with an accuracy higher than 0.3 pixels or 1 mas. This is a challenging task for images with saturated PSF or frames taken with a coronagraph. 

This can be solved by switching between short unsaturated cycles in FP mode for the beam shift measurement and long, peak-saturated disk observations taken in SP mode. The position of the star on the detector and the beam shift change only slowly with altitude and parallactic angle. Therefore they can be interpolated, for instance, as a function of time, using FP cycles with well-defined intensity peaks. These interpolations can be applied to the saturated images of the SP cycles.

Following this strategy, we recorded a total of six blocks of the FP cycles alternating with four blocks of the SP cycles on 2018 February 28, and two blocks of the FP cycles, one before and one after an SP block on 2018 June 22. Each cycle consisted of four consecutive measurements with different HWP offset angles of 0$^{\circ}$, 45$^\circ$, 22.5$^\circ$, and 67.5$^\circ$ switching the Stokes parameters $+Q, -Q, +U,$ and $-U$, respectively.  

Half of the data from February and all data from June were taken with the sky field rotated on the detector by $60^{\circ}$ in order to better distinguish between the circumstellar polarimetric signal and noise. The DIT of one individual exposure is 1.1\,s in the FP mode and 10\,s in the SP mode. The total exposure time in each mode is given in Table~\ref{t_Settings}, including an overview of the observing conditions, which have a significant effect on the quality of the data, as shown in Appendix~\ref{s_QphiUphi_app}. The February observation started under a good seeing condition of 0.66$''$ and coherence time of 3.7~ms which constantly degraded and achieved $\sim$2$''$ and 2.1~ms, respectively, at the end of the observing run. The first data reduction showed a possible detection of the scattered light from a debris disk, but only in the data of the first polarimetric cycles. Therefore the measurements were repeated in June to improve the signal-to-noise ratio (S/N) of the data.

\section{Data reduction}\label{s_data}
\subsection{IRDIS and IFS datasets}
The IRDIS and IFS data were calibrated with the SPHERE Data Reduction and Handling (DRH) 
pipeline {\it esorex} \citep{Pavlov2008} and were processed at the SPHERE Data Center \citep{Delorme2017}. The calibration of raw data consisted of background subtraction, bad pixel correction, flat fielding, correction of the pixel distortion \citep{Maire2016}, and extraction of the IFS spectral data cube. Additional procedures for improving the wavelength calibration and correcting for the spectral cross-talk \citep{Mesa2015} were applied to the IFS data. 

The position of the star was determined by fitting a 2D Gaussian function to the four waffle spots in the ``center frame'' and determining the intersection point of lines connecting the centers of two opposite spots. These coordinates were used for recentering all frames in the data cubes.

The final calibrated datasets include two IRDIS temporal data cubes (K1 and K2 filters) and 39 IFS temporal data cubes (39 wavelength channels) with 48 frames each. The pixel scale of the IRDIS detector is 12.27 mas (K band), and a science frame is $1024 \times 1024$ pixels. In the IFS data the pixel scale is 7.46 mas and a science frame is $290 \times 290$ pixels.

\begin{figure*}   
\includegraphics[width=17cm]{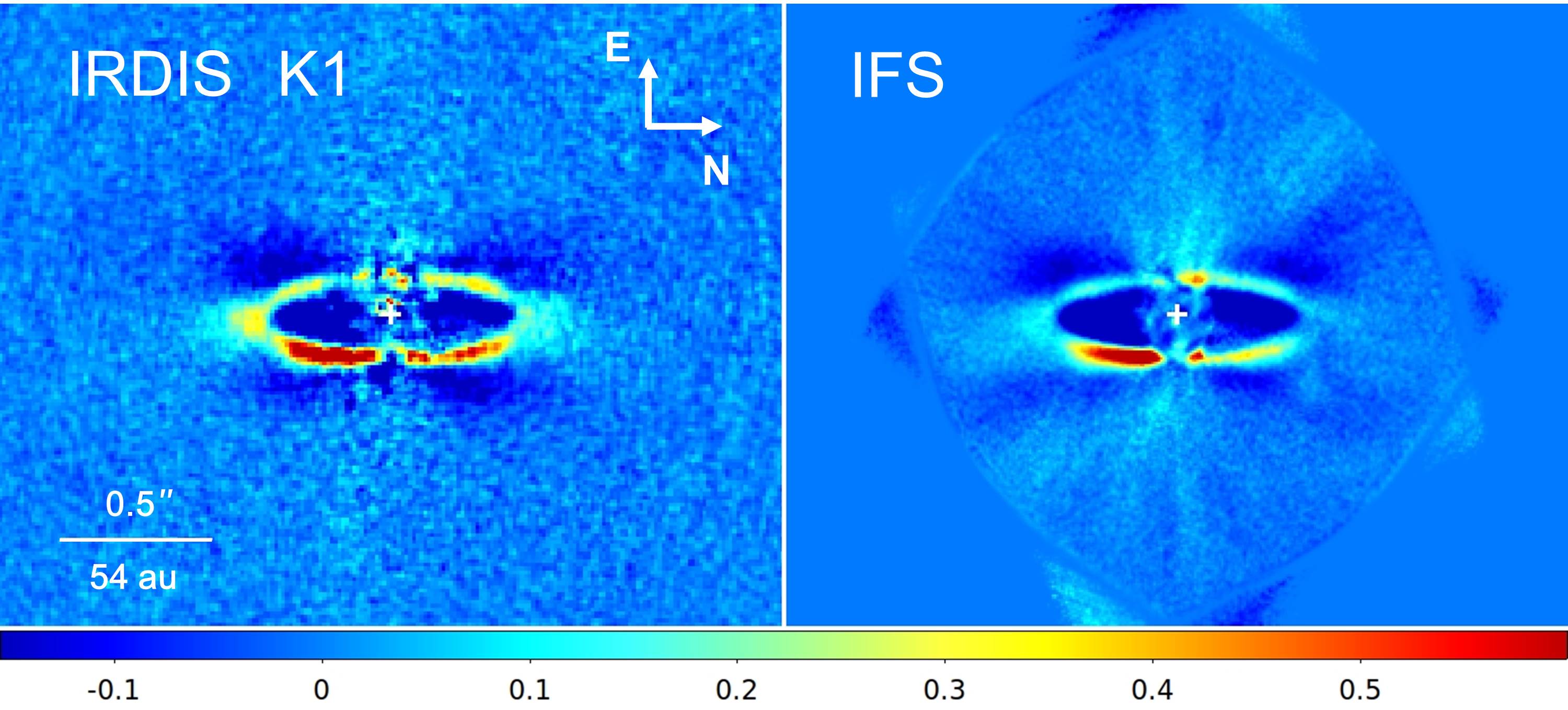} 
    \centering
    \caption{Total intensity images of the HD 117214 debris disk obtained with the cADI data reduction of the IRDIS K1 dataset (\textit{left panel}) and the spectrally combined IFS data (\textit{right panel}). The position of the star is marked by a white cross. The color bar shows the surface brightness in counts per pixel. \label{f_imaging}}
\end{figure*}

To subtract the stellar light, we used the SpeCal pipeline \citep{Galicher2018}, which provides several algorithms based on angular differential imaging \citep[ADI;][]{Marois2006} such as classical ADI (cADI), Principal Component Analysis \citep[PCA;][]{Soummer2012, Amara2012}, and Template Locally Optimized Combination of Images \citep[TLOCI;][]{Marois2014}, which were used to process both the IRDIS and IFS datasets.
Figure \ref{f_imaging} shows the cADI images of the K1 band data (left panel) and spectrally combined IFS data (right panel). Other data reductions are displayed in Figure~\ref{f_all_app}. 

\begin{figure*}   
    \includegraphics[width=17cm]{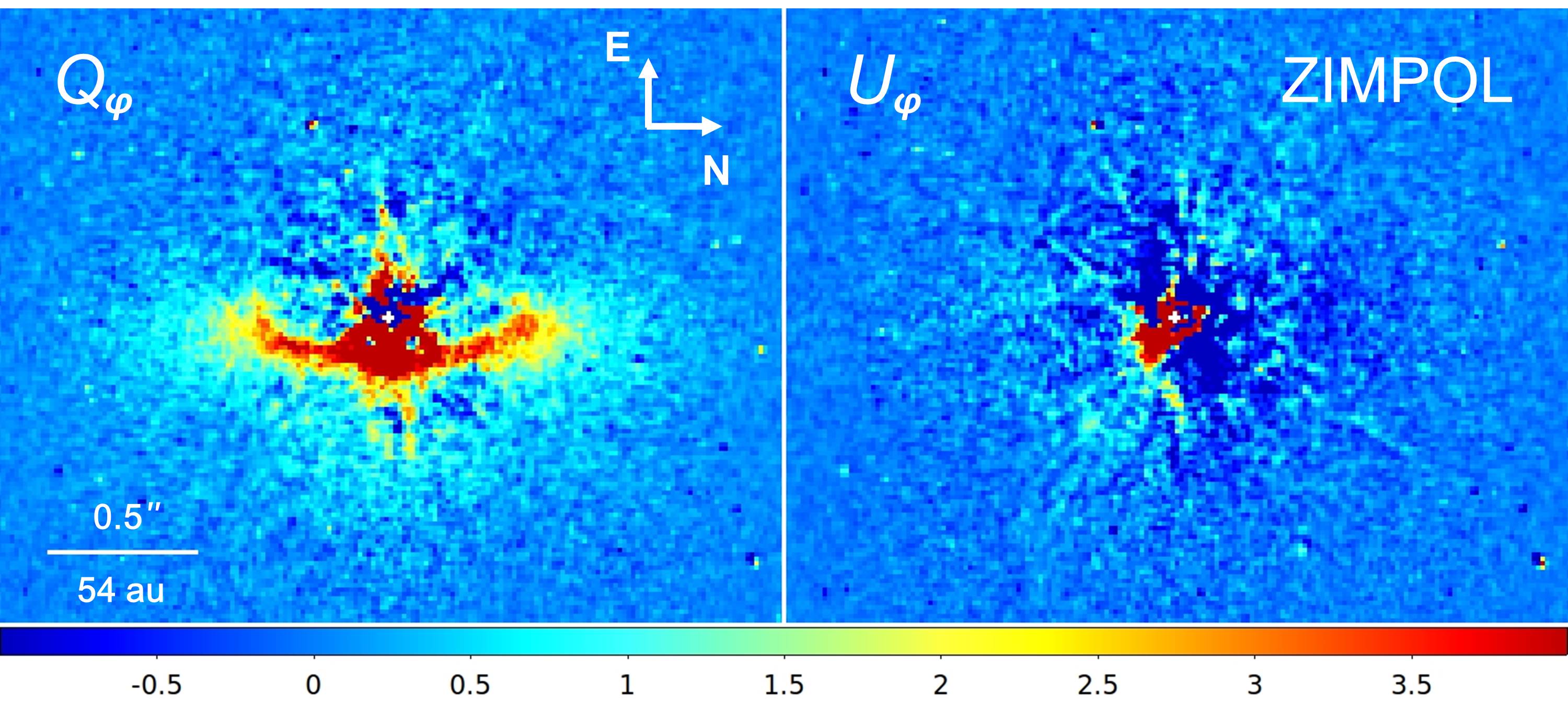} 
    \centering
    \caption{$Q_\varphi$ image (\textit{left panel}) showing the polarized intensity of scattered light and the $U_\varphi$ image (\textit{right panel}). The original data (OBS059\_0002-0049 and OBS173\_0001-0028, see Table~\ref{t_Settings}) were $4\times 4$ binned and smoothed with a Gaussian kernel with $\sigma_{\rm kernel} =$ 1 pixel to reduce the photon noise level. The position of the star is marked by a white cross. The color bar shows the surface brightness in counts per binned pixel. \label{f_QphiUphi}}
\end{figure*}

\subsection{ZIMPOL datasets}
The HD\,117214 data were reduced with the ZIMPOL data reduction 
pipeline developed at ETH Zurich. The pipeline includes preprocessing and calibration of the raw frames: subtraction of the bias and dark frames, flat-fielding, and correction for the modulation and demodulation efficiency. The instrumental polarization is corrected through the forced normalization of the fluxes in the $I_\perp$ and $I_\parallel$ frames as described in \citet{Engler2017}. 

To determine the beam offset between the $I_\perp$ and $I_\parallel$ frames and the position of the star in the combined intensity image, we fit a 2D Gaussian function to the stellar profile. This could only be applied to images with clean unsaturated PSF taken in the FP mode. To center the saturated science frames obtained in the SP mode, the position of the star on the detector and the beam offset were interpolated as a function of the local siderial time (see Sect.~\ref{s_Observations}) using the measurements of the FP cycles recorded immediately before and after the respective SP cycle.

\begin{figure*}
   \centering
  \includegraphics[width=16cm]{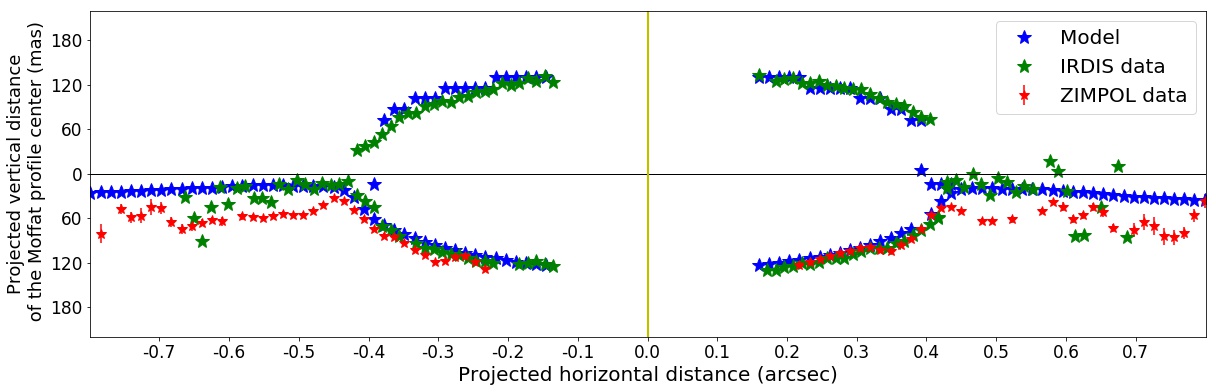} 
   \caption{Disk spine measured from the total intensity data (cADI, K1 filter) and polarized intensity data ($Q_\varphi$ image). The blue line shows the spine measured from model 2 (Col.~4, Table~\ref{t_results}). }  \label{f_spine}
\end{figure*}

The images of the Stokes parameter $Q$ and $U$ were calculated according to the double-difference method as follows:
\begin{equation}
Q = 0.5\cdot(Q^+ - Q^-)
\end{equation}
\begin{equation}
U = 0.5\cdot(U^+ - U^-)
,\end{equation}
and were then converted into the azimuthal Stokes parameters $Q_\varphi$ and $U_\varphi$ \citep[e.g., ][]{Engler2017}:
\begin{equation}
Q_\varphi = - Q\cos 2\varphi-U\sin 2\varphi
\end{equation}
\begin{equation}
U_\varphi = Q\sin 2\varphi - U\cos 2\varphi,
\end{equation}
where $\varphi$ is the polar angle measured EoN in the coordinate system centered on the star, and the sign convention for the $Q_\varphi = - Q_{\rm r}$ and $U_\varphi = - U_{\rm r}$ parameters defined in \citet{Schmid2006} was adopted.

The final format of the reduced images is $1024 \times 1024$ pixels; each pixel spans approximately $3.6 \times 3.6$ mas on sky. The full width at half-maximum (FWHM) of the stellar profile in the VBB is $\sim$7 pixels, corresponding to a resolution of $\sim$25 mas. To gain a higher S/N and preserve the spatial information, we applied a $4 \times 4$ binning to the polarized intensity data.

\section{Disk morphology}\label{s_Results}
\subsubsection*{Total intensity data} 
The IRDIS and IFS images (Fig.~\ref{f_imaging}) show a highly symmetric ellipse geometry without any observable center offset with respect to the star. The HD\,117214 disk appears to be a compact ring or belt with a radius smaller than 0.5$''$, an inclination of $70 - 75^\circ$, and a brighter side toward the west. The apparent difference in the surface brightness between the northern and southern side of the disk is most probably a result of the data post-processing. There seems to be a broad gap inside the ring between 0.4$''$ and at least 0.1$''$ (IWA of the coronagraph). 

\subsubsection*{Polarized intensity data}
Figure \ref{f_QphiUphi} shows the final $Q_\varphi$ and $U_\varphi$ images calculated as the mean of the data from the three best SP blocks: two first blocks recorded on 2018 February 28 (OBS059\_0002-0049, see Table~\ref{t_Settings}) with total $t_{\rm exp}= 32$ min, and one block from 2018 June 22 (OBS173\_0001-0028) with total $t_{\rm exp}= 16$ min. 

As expected for a circumstellar disk, the polarized scattered light is detected in the $Q_\varphi$ image (left panel in Fig.~\ref{f_QphiUphi}) and not in the $U_\varphi$ image (right panel in Fig.~\ref{f_QphiUphi}). The fainter side of the disk is also detected close to the northern belt ansae and is best seen in the $Q_\varphi$ image in the top row of Fig.~\ref{f_all_Qphi}. The signal of polarized light scattered off dust grains in the disk can be measured up to a distance of $\sim 1''$ from the star in our data.
 
\subsection{Position angle and spine of the disk}
The total intensity images (Fig.~\ref{f_imaging}) as well as 
the $Q_\varphi$ and $U_\varphi$ images displayed in Fig.~\ref{f_QphiUphi} were rotated first by 90$^\circ$ clockwise to place the disk axis horizontally. This includes a correction for the true north (TN) offset of instruments through the additional clockwise rotation by 1.75$^\circ$ \citep[IRDIS and IFS data,][]{Maire2016} or 2$^\circ$ (ZIMPOL data, Ginsky et al., in prep.). Figs.~\ref{f_imaging} and~\ref{f_QphiUphi} show that the disk major axis nearly coincides with the sky north-south axis, implying that the disk position angle (PA) is close to 180$^\circ$. To better determine the position of the disk major axis, we used the same method as described in \cite{Engler2018}: the total intensity images and the $Q_\varphi$ image were rotated stepwise within an interval of disk PAs between 175$^\circ$ and 185$^\circ$ by 0.5$^\circ$. At each step, the left half of the image was subtracted from the right half, and the residuals within the image area that contains disk flux were evaluated. The PA of the disk that corresponds to the residual minimum is equal to $180^{\circ} \pm 1^{\circ}$ (total intensity data) and $179^{\circ} \pm 1^{\circ}$ (polarimetry), including the TN offset. The disk major axis in the images in Figs.~\ref{f_imaging} and~\ref{f_QphiUphi} is placed at PA $=179.5^{\circ}$.

Figure~\ref{f_spine} shows the perpendicular offset from the disk major axis of the points with the highest flux as a function of the separation from the star along the major axis. The peak positions were found by fitting a Moffat function to the perpendicular profiles for the total intensity $I$ and polarized intensity $Q_\varphi$, as was done for the HIP 79977 debris disk \citep{Engler2017}. We call the curve that connects the profile peaks the spine of the disk.

The spine measured in the total intensity data (cADI, K1 filter) traces the ellipse of the dust belt that intersects the disk major axis at the radial distances of $r \approx 0.42''$ on both disk sides. At larger radial separations ($r > 0.42''$), the curve departs significantly from the major axis. The spine for $Q_\varphi$ coincides well with the intensity spine measured on the brighter disk side. Both datasets indicate no significant offset of the disk center with respect to the position of the star.

\begin{figure*}
\centering
\includegraphics[width=16.5cm]{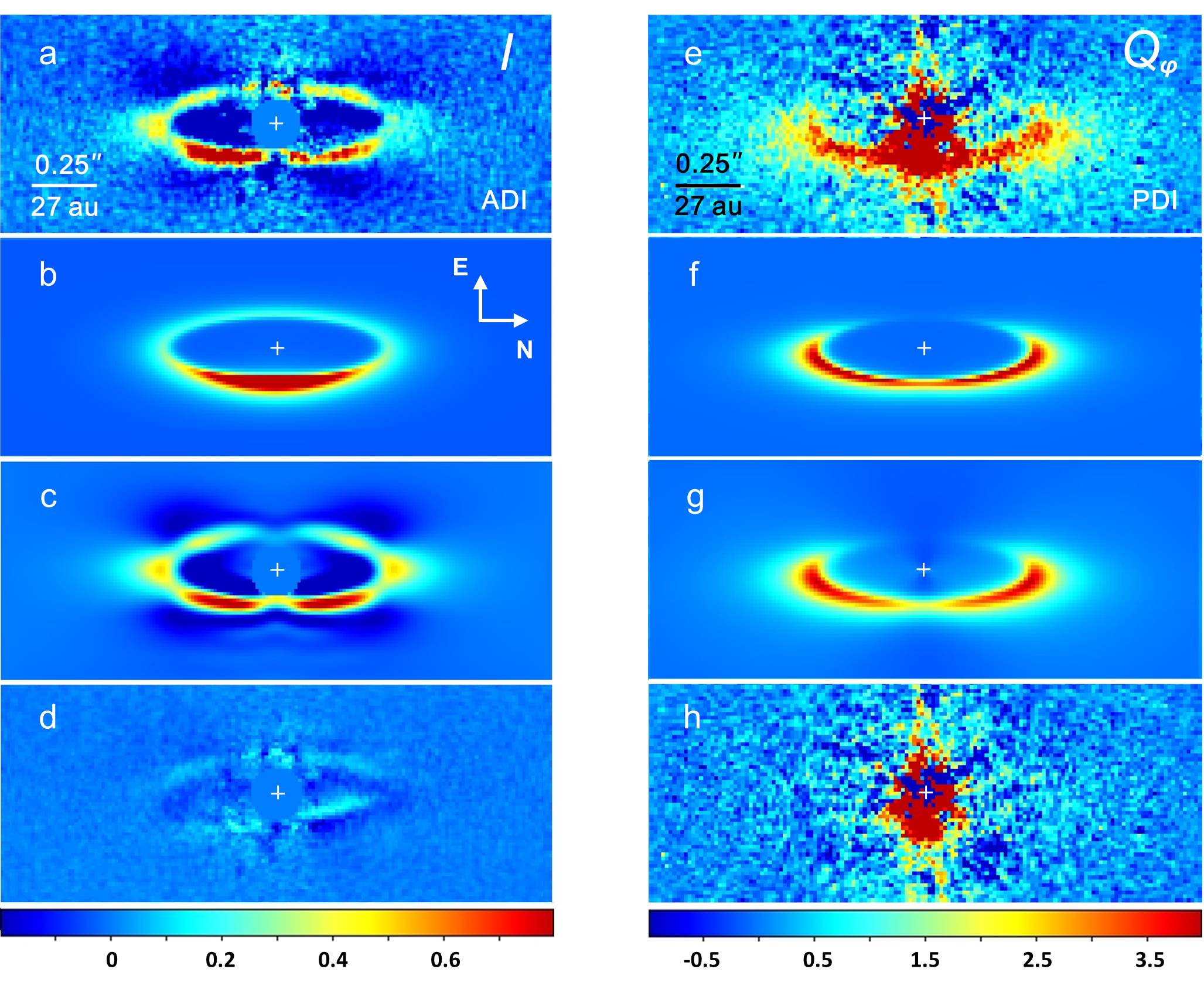} 
   \caption{Comparison of the total intensity image ({\it panel a}) and the $Q_\varphi$ image ({\it panel e}) with the models of scattered light ({\it panel b}) and scattered polarized light ({\it panel f}). {\it Panels c} and {\it g}: Models after convolution with the instrumental PSF and post-processing. {\it Panels d} and {\it h}: Residual images obtained after subtracting the model images from the total intensity image and the $Q_\varphi$ image. White crosses indicate the position of the star. Color bars show flux in counts per pixel.  \label{f_model}}
\end{figure*}

\subsection{Modeling the observed morphology} \label{Modelling}
\subsection*{Model with one Henyey-Greenstein parameter}
To model the observed disk surface brightness (left panel in Fig.~\ref{f_imaging}), we used the 3D single-scattering code presented in \citet{Engler2017}. We considered a rotationally symmetric distribution of dust grains in the disk, which can be described by the product of the radial power law and the Lorentzian profile for the vertical distribution of the grain number density \citep[e.g.,][]{Engler2018}. This model is based on the theory of the parent body belt that consists of massive planetesimals moving on a circular orbit with radius $r_0$. Mutual collisions between the planetesimals produce a large amount of micron-sized dust grains that are radially redistributed, and this is described by the radial power laws with the exponents 
$\alpha_{\rm in}> 0$ inside the belt and $\alpha_{\rm out}< 0$ for $r>r_0$. The Lorentzian profile is given by
\begin{equation}
f_L(h)=a_L{\left[ 1 + \left(\frac{h}{H(r)}\right)^2\right]}^{-1},
\end{equation}
where $h$ is the height above the disk midplane, and $a_L$ is 
the peak number density of grains in the disk midplane. 
The scale height of the disk $H(r)$ is defined as a half-width 
at half-maximum of the vertical profile at radial distance 
$r$ and scales like $H(r)=H_0\,(r/r_0)\,^\beta$, where $H_0= H(r_0)$ and $\beta$ is the disk flare index. 

The amount of light scattered into the line of sight depends on the scattering angle $\theta$. This dependence is described by the phase function $f_{\rm sca}$, which is often approximated by the Henyey-Greenstein (HG) function \citep{Henyey1941}:
\begin{equation} \label{e_HG}
f_{\rm sca}(\theta)=\frac{1-g^2}{4 \pi (1+g^2-2 g cos (\theta))^{3/2}},
\end{equation}
with $g$ as the HG scattering asymmetry parameter ($-1\leqslant g \leqslant 1 $). 

We assumed that dust grains scatter more radiation in the forward direction. This means that the asymmetry parameter $g$ has a positive value and the brighter side of the disk is closer to the observer. Furthermore, we assumed that the dust grains have the same properties everywhere in the disk and that scattered-light images preferentially trace dust grains with a size comparable to the wavelength of observation. This implies that the average scattering cross-section per particle is constant throughout the disk and can be included as a free parameter into the scaling factor $A$ of the model.

All model parameters are listed in Table~\ref{t_results}. To constrain them, we ran the custom Markov chain Monte Carlo (MCMC) code using the Python package {\it emcee} by \cite{Foreman-Mackey2013} and fit the synthetic model images of scattered light to the total intensity image in the K1 band (Fig.~\ref{f_model}a). At each MCMC step we tried a new model with a parameter set drawn from prior distributions, convolved the model with the instrumental PSF, and inserted it into an empty data cube at different position angles to mimic the rotation of the sky field during the observation. A cADI forward modeling was then performed to compare the result with the imaging data (see also Appendix~\ref{s_modeling_app}).

The obtained posterior distribution of the parameters are shown in Fig.~\ref{f_mcmc_1g}, and their median values, our best-fit parameters, are listed in Col.~3 of Table~\ref{t_results}. We refer to the disk model with these parameters as model 1 and obtain a reduced $\chi^2_\nu = 1.69$ for it with the degree of freedom $\nu = 4475$ (the degree of freedom is equal to the number of pixels minus the number of free parameters).

Fig.~\ref{f_mcmc_1g} shows that the MCMC converges to the solution for the radius of the planetesimal belt of $0.41''$ ($\sim$45 au), inclination of $\sim$70$^\circ$, and HG asymmetry parameter of 0.33. There is a negative correlation between the index of the outer power law for the radial distribution of grain number density $\alpha_{\rm out}$ and the flare index of the disk $\beta$. Both parameters seem to be constrained with the best-fitting values $\alpha_{\rm out} \approx -5$ and $\beta \approx 1.5$. The inner radial index $\alpha_{\rm in}$ has a relatively high value of 20, indicating a sharp inner edge of the debris belt. The disk scale height $H_0$ is estimated to be $\sim$0.0015$''$: this yields the aspect ratio between the disk radius and height of 0.004. This value is much lower than the minimum aspect ratio of 0.04 predicted by \citet{Thebault2009} for the observations of debris disks at visual and mid-IR wavelengths.

\begin{table*}  
  \caption[]{Disk model parameters. \label{t_results}}
             \centering
         \begin{tabular}{lccc}
           \hline
            \hline
            \noalign{\smallskip}
           \multirow{ 2}{*}{Optimized parameter} &  \multirow{ 2}{2 cm}{Priors} & Model 1 ($\chi^2_\nu = 1.69$ ) & Model 2 ($\chi^2_\nu = 1.56$ )  \\
            & & 1 HG parameter & 2 HG parameters \\
            \hline
            \hline
            \noalign{\smallskip}
       Radius of the belt $r_0$ ($''$ (au)) &[0.35, 0.55] & 0.41$^{+0.01}_{-0.01}$ (44.2$^{+1.1}_{-1.1})$ & 0.42$^{+0.01}_{-0.01}$ (45.2$^{+1.1}_{-1.1}$) \\[5pt]
       
       Scale height $H_0$ ($''$ (au)) &[0.0, 0.1] & 0.002$^{+0.001}_{-0.001}$ (0.2$^{+0.1}_{-0.1}$) & 0.005$^{+0.003}_{-0.002}$ (0.5$^{+0.3}_{-0.2}$) \\[5pt]
       
           Inner radial index $\alpha_{\rm in}$ & [0, 50] & 20$^{+7}_{-5}$ & 24$^{+18}_{-10}$ \\[5pt]
           
           Outer radial index $\alpha_{\rm out}$ & [-15, 0] & -5$^{+1.2}_{-1.5}$ & -4.2$^{+0.3}_{-0.5}$ \\[5pt]
           
           Flare index $\beta$ &[0, 6] & 1.5$^{+0.7}_{-0.7}$  & 0.3$^{+0.5}_{-0.3}$  \\[5pt]
           
           Inclination $i$ ($^\circ$) &[65, 85] & 71.0$^{+1.3}_{-1.0}$ & 73.2$^{+0.5}_{-0.5}$  \\[5pt]
           
           Position angle ($^\circ$) &[170, 190] & 179.0$^{+0.3}_{-0.0}$ & 179.4$^{+0.2}_{-0.2}$ \\[5pt]
           
           HG parameter $g_{1}$ &[0, 0.9] & 0.33$^{+0.01}_{-0.01}$ & 0.61$^{+0.02}_{-0.02}$  \\[5pt]
           
           HG parameter $g_{2}$ &[-0.4, 0.5] & (...) & -0.22$^{+0.04}_{-0.04}$  \\[5pt]
           
           Scaling parameter $w$ &[0, 1] & (...) & 0.73$^{+0.02}_{-0.02}$  \\[5pt]
           
           Scaling factor $A_p$ & [0, 100] & 25$^{+5}_{-9}$ & 11$^{+3}_{-2}$  \\[5pt]
           \noalign{\smallskip}
           \hline
           \hline
            \noalign{\smallskip}
      \end{tabular}
\end{table*}

\subsection*{Model with two Henyey-Greenstein parameters}
When it is subtracted from the total intensity image in K1 band (Fig.~\ref{f_model}a), model 1 (Col.~3 of Table~\ref{t_results}) leaves nonmarginal residuals on the west side of the disk (Fig.~\ref{f_residuals}b). Therefore we tested a second model with another phase function given by a linear combination of two HG scattering functions \citep[][]{Engler2017}:
\begin{equation} \label{eq:phase func}
f(\theta,g_1,g_2)= w \cdot f(\theta,g_1) + (1-w) \cdot f(\theta,g_2)\, ,
\end{equation} 
where the first parameter $g_1$ describes a strong diffraction peak, the second parameter $g_2$ represents the more isotropic part of the SPF, and $w$ is the scaling parameter, $0\leq w\leq 1$.

The posterior distributions of the fitted parameters derived with MCMC using the second model are shown in Fig.~\ref{f_mcmc_2g}, and their median values are given in Col.~4 of Table~\ref{t_results}. We consider this parameter set as a best fit to the data and used it to create a model denoted model 2. The values for the radius and inclination of the belt as well as the indexes of the outer and inner power laws for the radial distribution of the grain number density of model 2 are consistent with those of model 1. The slightly lower outer power-law index $\alpha_{\rm out} = -4$ of model 2 leads to a higher scale height of the disk ($H_0= 0.004''$) and a lower flare index ($\beta =0$) than for model 1 because of the degeneracy between these parameters, which is also shown in Fig.~\ref{f_mcmc_1g}. 

It is interesting to note that the MCMC favors an asymmetry parameter of $g_{1} = 0.66$ in combination with a negative parameter of $g_{2}= -0.22$ (25\% contribution). This corresponds to a strongly forward- and a slightly backward-scattering behavior of dust grains. Similar phase functions have been measured for the debris disk HD 35841 \citep{Esposito2018} and coma dust of comet 67P/Churyumov–Gerasimenko \citep{Bertini2017} in the Solar System (see Sect.~\ref{s_SPF}). 

\begin{figure}
\centering
\includegraphics[height=7.5cm]{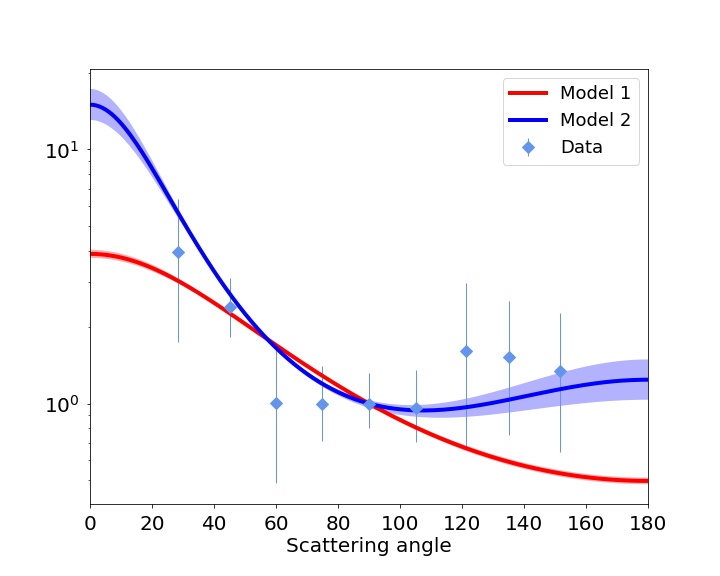}
   \caption{Comparison of the SPFs with one HG function (model 1 in Table~\ref{t_results}) and a combination with two HG functions (model 2 in Table~\ref{t_results}). Blue diamonds show the measured phase function in the total intensity disk image. Red and blue shaded areas cover the range of uncertainty on parameters obtained from the MCMC posterior distributions. \label{f_model_SPF}}
\end{figure}

To compare this combination of asymmetry parameters with the HG function obtained for model 1, we plot both SPFs in Fig.~\ref{f_model_SPF}. The disk flux measured from the total intensity image in K1 band (left panel of Fig.~\ref{f_imaging}) is also shown by blue diamonds in this plot. The flux was integrated in circular apertures with a radius of 0.04$''$ placed along the disk spine at different scattering angles. The data in Fig.~\ref{f_model_SPF} show the average value of the north and south disk sides multiplied by a correction factor to account for the flux loss caused by the ADI data post-processing. The mean correction factor was estimated with forward-modeling by comparing between flux values measured in the same apertures before and after the ADI processing of models. Both phase functions, with one and two HG parameters, as well as the disk flux are normalized to their values at scattering angle of 90$^\circ$. We took into account that the pixel noise might be correlated and calculated the uncertainty on flux as the sum of flux errors for the individual pixels within each aperture. 

The flux error bars are relatively large because of the small angular size of the HD 117214 debris disk. It is one of the smallest disks that have been resolved in scattered light so far. Based on the visual comparison (Fig.~\ref{f_model_SPF}), it is therefore not possible to determine which phase function fits the data better. Moreover, because the disk has an inclination of $\sim$70$^\circ$, scattering angles smaller than 20$^\circ$ and larger than 160$^\circ$ are not accessible for the observations. The phase function in the entire range of scattering angles might therefore show a stronger (or weaker) forward- and backward-scattering behavior than our result.

In order to asses the goodness of fit of models 1 and~2, we computed the reduced $\chi^2_\nu$ (for model 2 the degree of freedom $\nu$ is equal to 4473) for both models and obtained $\chi^2_\nu = 1.69$ and $\chi^2_\nu = 1.56$, respectively. According to the $\chi^2$ criteria, this indicates that model 2 fits the data better than model~1. The visual comparison of the two residual images (Fig.~\ref{f_residuals}) also confirms that the model with two HG parameters leaves lower residuals after the model image is subtracted from the data. Therefore we show model 2 in Fig.~\ref{f_model}b and plot the vertical offset of the spine of this model in Fig.~\ref{f_spine} (blue asterisks) for comparison with the data. The spine offset of the model 2 was measured by fitting the Moffat profiles to the disk cross-sections, as was done with the total intensity data in K1 band (Sect.~\ref{s_Results}). 

Using the parameter set of model 2, we also created a model of the disk polarized intensity shown in Fig.~\ref{f_model}f and Fig.~\ref{f_model}g in comparison to the $Q_\varphi$ image (Fig.~\ref{f_model}e). As the polarized phase function we used the product of the scattering function $f(\theta,g_1,g_2)$ (Eq.~\ref{eq:phase func}) with the polarization fraction \citep{Engler2017}
\begin{equation} \label{e_pol}
f_{\rm pol}(\theta)= f(\theta,g_1,g_2) \cdot p_m\frac{1- \cos^2\theta}{1+\cos^2 \theta}
,\end{equation}
where $p_m$ is the maximum fractional polarization at a scattering angle of $\theta=90^\circ$ ($0 \leqslant p_m  \leqslant 1 $). 

Based on the residual image (Fig.~\ref{f_model}h), we consider that the model image of the polarized light (Fig.~\ref{f_model}g) matches the $Q_\varphi$ image (Fig.~\ref{f_model}e) well. Therefore the same model was used to estimate the effect of the polarized flux cancellation (Sect.~\ref{s_Photometry}).

\section{ZIMPOL photometry}\label{s_Photometry}
In order to compare the stellar flux with the amount of scattered light, we considered the ZIMPOL dataset. The polarimetric measurement does not require the ADI technique, and therefore the ZIMPOL data do not suffer from the self-subtraction effect, which is difficult to quantify even by forward-modeling.

\subsection{Stellar flux} \label{s_Star_flux}
To obtain the stellar count rate in the VBB filter, we used unsaturated intensity images $I_\perp$ and $I_\parallel$ taken during the FP cycles. The stellar intensity was calculated as a mean of the sum of the intensities measured for both polarization states in the $Q$ and $U$ cycles,
\begin{equation} \label{e_I}
I = \frac{I^Q_\perp + I^Q_\parallel + I^U_\perp + I^U_\parallel}{2}.
\end{equation}
The mean count rate for the central star of $(4.26 \pm 0.16)\times 10^6$ counts per second (ct/s) per ZIMPOL arm was obtained by summing all counts registered within the circular aperture with a radius of $1.5''$ (413 pixels). This count rate yields a photometric magnitude in the VBB of $m(\mathrm{VBB})$ = 7.72$^m \pm 0.06^m$ according to \citet{Schmid2017},
\begin{equation} \label{e_magn}
m(\mathrm{VBB}) = -2.5 \log (\mathrm{ct/s})-\mathrm{am}\cdot k_1(\mathrm{VBB})-m_{\mathrm{mode}}+ \mathit{z}p_{\mathrm{ima}}(\mathrm{VBB}),
\end{equation}
where am~$=1.25$ is the airmass, $ k_1(\mathrm{VBB})=0.086^m$ is 
the filter coefficient for the atmospheric extinction, 
$\mathit{z}p_{\mathrm{ima}}(\mathrm{VBB})=24.61^m$ is the photometric 
zero-point for the VBB filter, and $m_{\mathrm{mode}}= 0.18^m$ is an offset 
to the zero-point accounting for the fast polarimetry detector mode. 
The derived magnitude in the VBB agrees well with the literature values for the HD\,117214 photometric magnitudes at adjacent wavelengths (see Table~\ref{t_filters}). 

\begin{table}  
      \caption[]{HD 117214 photometry.  \label{t_filters} }
          \centering
             \begin{tabular}{lccccc}
            \hline
            \hline
            \noalign{\smallskip}
            Filter & $\lambda$ & $\Delta \lambda$ & mag & $\sigma_{\rm mag}$ & Ref. \\
                   & ($\mu$m) & ($\mu$m) & (mag) &  (mag) &  \\
            \hline
            \noalign{\smallskip}
            Tycho V & 0.532 & 0.095  & 8.06 & 0.01 & 2 \\
            Gaia G  & 0.673 & 0.440 & 7.96 & <0.01 &  3\\
            ZIMPOL VBB  & 0.735 & 0.290 & 7.72 & 0.06 &4  \\  
            2MASS J & 1.250 & 0.300 & 7.17 & 0.02 &5  \\
            2MASS H & 1.650 & 0.300 & 6.97 & 0.04 &5  \\
  \noalign{\smallskip}
            \hline
            \noalign{\smallskip}
         \end{tabular}
\tablebib{ (1)~\citet{ESA1997}; (2) \citet{Hog2000}; (3) \citet{GaiaCollaboration2018}; (4) this work;
(5) \citet{Cutri2003}. }         
  \end{table}

\subsection{Polarized flux and contrast of the disk} \label{s_Disk_flux}
We measured the polarized flux of the disk by summing all the counts in two rectangular areas with a height of $\sim$0.51$''$ and a width of $\sim$0.92$''$, which enclose the regions with the radial separations $0.19'' < r < 1.10''$ that lie below the disk major axis up to $\sim$0.35$''$ and above the disk axis up to $\sim$0.16$''$. The total mean count rate within these areas is $3165 \pm 260$ ct/s per ZIMPOL arm in the SP mode. The actual polarized flux from the disk is expected to be higher because the measuring areas only cover a part of the disk. The measured flux is also reduced because of the polarimetric flux cancellation caused by the instrumental PSF \citep[e.g.,][]{Engler2018}. With the model presented in Sect.~\ref{Modelling}, we took these effects into account and estimate that the total intrinsic polarized flux is higher by a factor of three than the measured flux and amounts to $9600 \pm 800$ ct/s. This  count rate corresponds to the disk magnitude of $mp_{\mathrm{disk}}(\mathrm{VBB})$ = 16.48$^m \pm$ 0.3$^m$ in the VBB filter according to Eq.~\ref{e_magn} when a throughput offset parameter $m_{\mathrm{mode}}= -1.93^m$ is used for the SP mode. 
The estimated magnitudes of the star and disk yield a ratio between the disk total polarized flux and 
stellar flux of $(F_{\rm pol})_{\rm disk}/F_{\rm \ast} = (3.1 \pm 1.2)\cdot 10^{-4}$ or the disk polarized flux contrast of $mp_{\mathrm{disk}}(\mathrm{VBB}) - m \mathrm{_{star}(VBB)} = 8.76$ mag. 

The maximum surface brightness per binned pixel (0.015$'' \times 0.015''$) of $\sim$7~ct/s is measured in the north of the star in the bright region at a radial separation of $r\approx 0.38''$. This peak corresponds to the magnitude ${\rm SB} \mathrm{_{peak}(VBB)} = 15.1^m \pm 0.3^m$ arcsec$^{-2}$ or surface brightness contrast for the polarized flux of ${\rm SB} \mathrm{_{peak}(VBB)} - m \mathrm{_{star}(VBB)} = 7.4$ mag arcsec$^{-2}$.

\section{Discussion} \label{Discussion}
\subsection{Constraints on dust properties}
In this Section, we would like to discuss how our results can be used to place some constraints on properties of dust grains in the debris disk HD 117214. In particular, this can be done from comparison between the total polarized flux and thermal emission of the disk (Sect.~\ref{s_Lambda parameter}), or from comparison of the measured HD 117214 SPF with the phase functions obtained for various dust populations (Sect.~\ref{s_SPF}). 

\subsubsection{Comparison between the polarized flux and thermal emission} \label{s_Lambda parameter}

\subsubsection*{Comparison with the debris disk HIP 79977} 
By its geometric and photometric characteristics, the HD 117214 disk is very similar to the debris disk around another F star in the Sco-Cen association:  HIP 79977 (HD 146897). In both disks (see Table~\ref{t_two_disks}), the parent body ring has a radius larger than 40 au, and most of the debris material is located well beyond the snow line and forms a so-called extrasolar Kuiper belt. 

The fractional IR excess of HD 117214 $L_{\mathrm{IR}}/L_{\ast}=2.53\times 10^{-3}$ 
is only half as high as that of the HIP 79977 disk \citep{Jang-Condell2015}, indicating a less massive disk.
The optical properties of the dust grains in both disks, however, might be very similar. \cite{Engler2017} introduced a $\Lambda$ parameter to characterize the ratio between the fractional scattered polarized flux $(F_{\rm pol})_{\rm disk}/F_{\rm \ast}$ (Sect.~\ref{s_Disk_flux}) and the fractional IR luminosity of the disk $L_{\mathrm{IR}}/L_{\ast}$. For the HIP 79977 disk we obtained $\Lambda_{\rm \,HIP\,79977} = 0.11 \pm 0.02$. The $\Lambda$ parameter of the HD 117214 disk is the same within uncertainties:
\begin{equation*}
\Lambda_{\rm \,HD\,117214} = {(F_{\rm pol})_{\rm disk}/F_{\rm \ast} \over L_{\mathrm{IR}}/L_{\ast}}= \frac{(3.1 \pm 1.2)\cdot 10^{-4}}{2.53\cdot 10^{-3}} = 0.12 \pm 0.05.
\end{equation*}

\subsubsection*{Scattering albedo of the disk} 
As shown by \cite{Engler2017}, the flux ratio $(F_{\rm pol})_{\rm disk}/F_{\rm \ast}$ can be used as a good order-of-magnitude estimate for the fractional polarized light luminosity of the disk $(L_{\rm pol})_{\rm disk}/L_{\rm \ast}$ because this ratio is obtained in the VBB filter near the peak of the stellar energy distribution. Thus the measured polarized flux $(F_{\rm pol})_{\rm disk}$ allows us to estimate the scattered-light luminosity of the disk $L_{\rm sca}$ by calculating the ratio $(F_{\rm pol})_{\rm disk}/L_{\rm sca}$ from the model \citep{Engler2017}. Generally, this ratio depends on the wavelength, asymmetry parameter $g,$ and inclination of the disk. When the wavelength dependence of the dust scattering is neglected, the relation between $(F_{\rm pol})_{\rm disk}$ and $L_{\rm sca}$ (per steradian) for a disk with the best-fitting parameters (Col.~4, Table~\ref{t_results}) is given by
\begin{equation}
L_{\rm sca} = \frac{1}{4 \pi} \frac{(F_{\rm pol})_{\rm disk}}{0.03\, p_m}= \frac{(F_{\rm pol})_{\rm disk}}{0.38\, p_m}
,\end{equation}
where the factor of 0.03 is obtained for model~2.
We estimate that the uncertainty for this relationship resulting from our assumptions on the SPF is below 15\%.

The reflectivity of dust in the HD 117214 disk can be characterized by comparing the fractional scattered-light luminosity with the fractional IR excess luminosity. For this purpose, we defined the scattering albedo of the disk as a relation between the amount of stellar radiation scattered by dust grains to the amount of the radiation attenuation due to dust scattering and absorption, where the latter is represented by the fractional IR excess luminosity of the disk,
\begin{equation} \label{e_albedo}
\begin{split}
\omega_{\rm \,HD\,117214} = {L_{\rm sca}/L_{\rm \ast} \over {L_{\mathrm{IR}}/L_{\ast} + L_{\rm sca}/L_{\rm \ast} } }= \frac{1}{1+ {L_{\mathrm{IR}}/L_{\ast} \over L_{\rm sca}/L_{\rm \ast}}}  = \\ 
= \frac{1}{1+ \frac{0.38\, p_m} {0.12}} = \frac{1}{1+ 3.17\,p_m}
\end{split}
.\end{equation}

We cannot measure the total intensity of scattered light in the ZIMPOL VBB filter, and therefore we are unable to calculate the maximum polarization fraction of scattered light for the HD 117214 disk.  
For some other debris disks, the fractional polarization was measured and is in the range between 10\%, for example, for $\beta$ Pic \citep{Tamura2006} or HD 32297 \citep{Asensio-Torres2016} and 40\%, for example, for HR 4796 A \citep{Milli2019}. 

\begin{table}  
      \caption[]{Comparison of stellar and debris disk properties between HD 117214 and HIP 79977.  \label{t_two_disks} }
          \centering
             \begin{tabular}{lccc}
            \hline
            \hline
            \noalign{\smallskip}
            Parameter & HD 117214 & HIP 79977 & Ref. \\
             \hline
            \noalign{\smallskip}
                    Spectral type & F6V & F2/3V & 1, 2 \\ 
            Stellar luminosity ($L_\odot$) & 5.64 & 3.66 & 3 \\
            Stellar mass ($M_\odot$) & 1.6 & 1.5 &  3 \\        
            Disk IR excess  & $2.53\times 10^{-3}$ & $5.21\times 10^{-3}$ & 3\\
            Disk radius (au) & 45 & 53--73 & 4, 5, 6  \\           
            Minimum grain size\tablefootmark{a} ($\mu$m)& 1.5 & 1 & 3  \\

           \noalign{\smallskip}
            \hline
            \noalign{\smallskip}
         \end{tabular}
\tablebib{
(1)~\citet{Houk1975}; (2) \citet{Houk1988}; (3) \citet{Jang-Condell2015}; (4) this work;
(5) \citet{Goebel2018}; (6) \citet{Engler2017}. 
}
\tablefoot{
\tablefoottext{a}{The quoted minimum grain size was estimated by \cite{Jang-Condell2015} assuming that 1) the dust is composed of amorphous olivine and pyroxene, and 2) the smallest dust grains are removed by radiation pressure from the system.}
}
 \end{table}  
   
It would be very useful to determine the maximum polarization fraction and thus the grain albedo. Both characteristics contain an information about particle composition, size, and shape \citep[e.g.,][]{Graham2007, Choquet2018}. In particular, the scattering albedo describes the properties of the grain surface. Grains with a black surface ($\omega < 0.1$) absorb all incident radiation. In contrast, bright grains ($\omega$ close to 1) reflect most of the incident light. A high albedo ($> 90$\%) might indicate pure icy grains, whereas an albedo one order of magnitude lower would suggest, for instance, dirty water-ice grains with inclusions of dark material such as carbon \citep[e.g.,][]{Mukai1986, Preibisch1993}.

Additionally, the multiband observations performed with narrow filters would help not only to determine the disk color, but also to probe the spectral albedo. Assuming that the scattered light images trace the population of dust grains with sizes similar to the wavelength of observation, the spectral albedo would allow us to investigate the differences in scattering behavior between the grains of different sizes, and, finally, to conclude about the amount of their contribution to the thermal emission of the debris disk.

\begin{figure*}
\centering
\includegraphics[width=18cm]{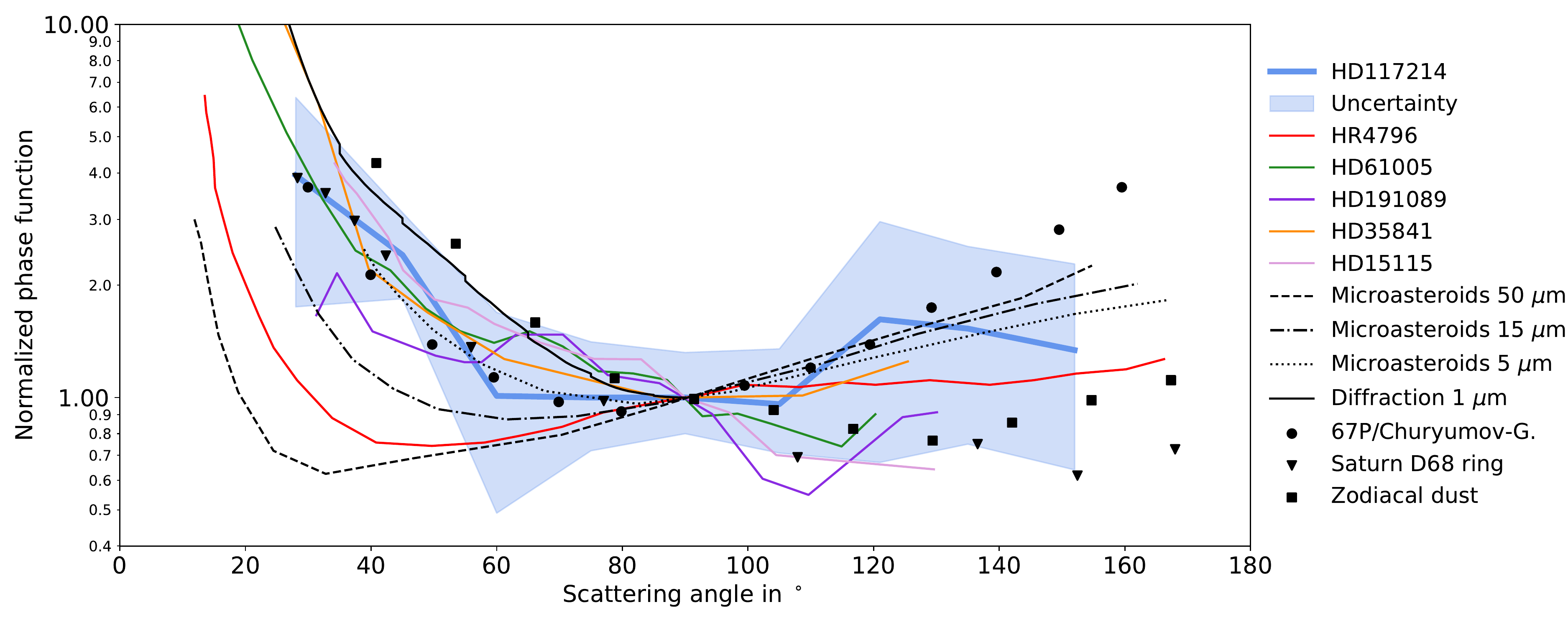}
 \caption{Observed SPFs measured for different debris disks and dust populations in the Solar System. The black lines show the micro-asteroid SPFs derived by \citet{Min2010} and the diffraction model for micron-sized grains. For the references of data used in this figure, see Table~\ref{t_SPF_obs}. \label{f_SPFs}}
\end{figure*}

\subsubsection{Scattering phase function} \label{s_SPF}
In addition to the scattering albedo, the phase function (see Sect.~\ref{Modelling}) can provide information on dust properties as well. The shape of the SPF depends on various parameters such as size, composition, or the shape of dust grains as well as on the observation wavelength. Therefore, there is a great interest to retrieve these dust characteristics by fitting the empirical SPF with theoretical phase functions, which can be derived, for instance, using the Mie theory, Fraunhofer diffraction, T-matrix  approximation \citep{Mishchenko1996}, or the Hapke reflectance \citep{Hapke1981}.

Laboratory experiments with numerous dust mixtures \citep[e.g.,][]{Pommerol2019, Frattin2019, Munoz2017, Poch2016} and in situ measurements of the scattered flux from different dust populations in the Solar System \citep[e.g.,][]{Hedman2015, Bertini2017} have shown that the theoretical phase functions can match the data well \citep[e.g.,][]{Moreno2018}. 

However, it is not straightforward to interpret the fitting results in terms of particle composition or even a power-law index for the grain size distribution \citep[e.g.,][]{Hedman2015, Milli2017}, and it remains often ambiguous for various reasons. On the one hand, model fitting requires very accurate measurements, which are particularly difficult to obtain when debris disks of distant stellar systems are observed. The flux obtained from the total intensity image of the disk should be corrected for (1) disk self-subtraction caused by the post-processing technique, (2) the projection effects due to disk inclination, and (3) the blurring effect of the PSF. These corrections are model-dependent and introduce a large uncertainty on the measured value. On the other hand, many different dust mixtures provide a similar phase function. Therefore, the SPF fitting to the limited range of scattering angles does not allow a conclusive statement. 

This can be the case even if the uncertainty of flux measurement is small, as the example of two dusty rings of Saturn shows. \cite{Hedman2015} estimated the brightness of the G ring and the D68 ringlet in a wide scattering angle range from $0.5^\circ$ to $170^\circ$ using the data taken by the narrow angle camera (NAC) and wide angle camera (WAC) of the Imaging Science Subsystem (ISS) on board the {\it Cassini} spacecraft in 2006. The Mie-based calculations to reproduce the SPFs of both rings did not provide strong constraints on particle composition. The best-fit solutions prefer relatively low fractions of water ice (< 30\% for the G ring and 10\% for the D68 ring), whereas the main rings of Saturn are expected to contain ice-rich particles based on measured strong water-ice absorption bands in the near-IR \citep{Cuzzi2009} and high reflectivity and low emissivity at radio wavelengths \citep{Pollack1975}.

In order to extract information about the particle size distribution, \citet{Hedman2015} fit the forward-scattering peaks of the observed SPFs of both rings with the Airy function predicted by Fraunhofer diffraction. The authors found that the particle size distribution does not follow a constant power law at small scattering angles. Instead, the power-law index for the differential size distribution significantly varies below $\theta=10^\circ$. A similar result was also obtained by \cite{Milli2017} for the debris disk HR 4796 A. \citet{Hedman2015} noted that in this case, the forward-scattering peak produced by a population of dust particles can be described by a Fraunhofer model for a diffraction of a single particle with a size $s$ equal to the average effective size $s_{\rm eff}$ of the dust population.

\begin{table}  
      \caption[]{References for data used in Fig.~\ref{f_SPFs}.  \label{t_SPF_obs} }
	  \centering
	     \begin{tabular}{lccccc}
            \hline
            \hline
            \noalign{\smallskip}
            Target & Instrument & Filter & $\lambda$ & Ref. \\
                   &   &   &  ($\mu$m)\tablefootmark{a}  &  \\
            \hline
            \noalign{\smallskip}
	   	    HD 117214 & SPHERE/IRDIS & K1 & 2.110 & 1 \\ 
            HR 4796 A & SPHERE/IRDIS & H & 1.626 & 2 \\ 
            HD 61005 & SPHERE/IRDIS & H & 1.626 & 3 \\ 
            HD 191089 & \textit{HST}/NICMOS & F110W & 1.100 & 4 \\ 
            HD 35841 & GPI & H & 1.647 & 5 \\ 
            HD 15115 & SPHERE/IRDIS & H & 1.626 & 6 \\ 
            Zodiacal & Rocket-borne & F1 & 0.476 & \multirow{ 2}{*}{7, 8} \\ 
             dust & photometer & F2 & 0.592 &  \\ 
            Saturn's D68 &  \multirow{ 2}{*}{\textit{Cassini}/ISS} & WAC clear & 0.634 & \multirow{ 2}{*}{9} \\ 
		    ring &  & NAC clear & 0.651 &  \\
		    \multirow{ 2}{*}{Comet 67P} & \multirow{ 2}{*}{\textit{Rosetta}/OSIRIS} & WAC F21 & 0.537 & \multirow{ 2}{*}{10} \\ 
		     &  & NAC F22 & 0.649 &  \\ 
	   \noalign{\smallskip}
            \hline
            \noalign{\smallskip}
         \end{tabular}
\tablebib{ (1) this work; (2) \cite{Milli2017}; (3) \cite{Olofsson2016}; (4) \cite{Ren2019}; (5) \cite{Esposito2018}; (6) \cite{Engler2019}; (7) \cite{Leinert1974}; (8) \cite{Leinert1976}; (9) \cite{Hedman2015}; (10)  \cite{Bertini2017}. }
\tablefoot{
\tablefoottext{a}{ Central or effective wavelength of filter. }
}
 \end{table}  
We did not achieve the same level of accuracy by measuring the SPF of the HD 117214 disk. Therefore we tried to gain insight about the possible composition or average size of the grains in this disk from comparisons between our measurement and the phase functions observed for some other debris disk systems and dust populations in the Solar System. In Fig.~\ref{f_SPFs} we plot the HD 117214 disk SPF (shown by blue diamonds in Fig.~\ref{f_model_SPF}) together with the empirical phase functions of debris disks HR 4796 A, HD 61005, HD 191089, HD 35841, HD 15115, the zodiacal dust and coma dust of comet 67P/Churyumov–Gerasimenko, and Saturn's ring D68 (see Table~\ref{t_SPF_obs} for references). The uncertainty of the measurements is shown only for the SPF of HD 117214 (this work) for clarity. 

The zodiacal dust is the interplanetary dust that originates mainly from the fragmentations of Jupiter-family comets \citep{Nesvorny2010} and collisions between asteroids in the asteroid belt \citep{Espy2006}. \citet{Leinert1976} derived the empirical SPF (Fig.~\ref{f_SPFs}) by combining several datasets of zodiacal dust photometry from different experiments at optical wavelengths. The SPF was calculated assuming that the spatial distribution of the grain number density in the inner Solar System scales like a radial power law $n\sim 1/r$.  

Comet 67P/Churyumov–Gerasimenko is a Jupiter-family comet with an orbital period of 6.45 years. The SPF of the dust from the cometary coma was measured in situ by the OSIRIS instrument on board the  \textit{Rosetta} spacecraft. Twelve multiwavelength series were acquired with the NAC and WAC to study the intensity of light scattered by the coma dust against the phase angle of the observations. The SPF represented in Fig.~\ref{f_SPFs} is from the dataset MTP020 recorded on 2015 August 28 at a heliocentric distance of 1.25 au (close to periastron) and at a distance to the comet nucleus of 420 km.  

In Fig.~\ref{f_SPFs} we also reproduce three theoretical SPFs derived by \citet{Min2010} for dust grains covered by small regolith particles, which reflect light backward like asteroidal bodies. The authors computed these micro-asteroid SPFs by applying Fraunhofer diffraction and Hapke reflectance \citep{Hapke1981} to the grains with radii of 5, 15, and 50 $\mu$m.\footnote{In the original paper by \cite{Min2010}, grain diameters are given in the notation instead of grain radii.} The diffraction part of the SPFs was averaged over a narrow flat size distribution to smooth the resonance structures. In the same way, we calculated a Fraunhofer diffraction of particles with a size of 1 $\mu$m, and this is also included in Fig.~\ref{f_SPFs} (black solid line). 

To facilitate comparison, all displayed SPFs were normalized to their values at scattering angle of 90$^\circ$. This yields for all SPFs, except for the HR 4796 A disk, a similar shape for scattering angles $\theta < 90^\circ$. This result, which was also found by \citet{Hughes2018}, is remarkable because the phase functions were derived from different dust populations residing in different environments. The apparent curve similarity between different objects of dust (excluding HR 4796 A) might indicate that one (or two) parameter is predominantly responsible for the shape of the forward-scattering peak, and that the dust populations discussed here probably has a similar value for this parameter. 

The plausible suggestion would be that an average effective size of particles plays an important role in this range of scattering angles.  Fig.~\ref{f_SPFs} shows that the diffraction curve we calculated for micron-sized particles reproduces the gradient of the phase function of the zodiacal light well. Modeling the observed brightness of the zodiacal light, \citet{Leinert1976} also found that the fine dust ($0.16 -29$ $\mu$m) with an average grain size of $\bar{s}=0.83$ $\mu$m contributes considerably (57\%) to the scattered-light intensity. The observed SPFs of coma dust in P67 and that of Saturn's ring match the diffraction contribution from grains with an average radius of 5-10 $\mu$m well. \citet{Hedman2015} therefore concluded that the typical particle size in the D68 ring of Saturn is about a few microns. 

If the peak of forward scattering from different types of debris is shaped mainly by the Fraunhofer diffraction, then the average effective grain size for the debris disks HD 117214, HD 61005, HD 191089, HD 35841, and HD 15115 is in the range between 1 and 10 microns, and between 15 and 50 $\mu$m for HR 4796 A. This is consistent with the minimum grain size of 18 $\mu$m derived by \citet{Milli2017} from the SPF fit using Mie theory. It is also consistent with the predicted cutoff of the grain size distribution caused by the radiation pressure of the host star, which removes the smallest grains from the system \citep{Burns1979}. HD 117214, HD 61005, HD 191089, HD 35841, and HD 15115 are all solar-type stars (from F4 to G8) with an expected blowout grain size of $0.8-1.5$ $\mu$m \citep[][Table~\ref{t_two_disks} of this work]{Olofsson2016}, while the estimated blowout grain size for the A0 star HR 4796 A is $\sim$10 $\mu$m \citep{Augereau1999}. 

If these estimates of the average effective grain size are valid, then the size parameter of particles $x = 2 \pi\, \bar{s}_{\rm eff}/ \lambda$ is close to or larger than $2 \pi$ in all presented observations (see Table~\ref{t_SPF_obs} for the effective or central wavelength of filters). Another implication is that the width of the forward-scattering peak should be different when the same target is observed in a significantly different wavelength range. 

Although all SPFs discussed here have a similar shape at scattering angles below $90^\circ$, in the range of $90^\circ < \theta < 180^\circ$ they show a noticeably different behavior. The phase function of the P96 coma dust exhibits strong backscattering and reaches similar values as at small scattering angles producing a u-shape, as is typical for comets \citep{Bertini2017}. In contrast, the SPF of the dust grains in the Saturn ring decreases. Some of the debris disk curves seem to have a positive slope, for instance, in HD 117241 and HD 35841. The other SPFs (HD 61005, HD 191089, and HD 15115) have a negative slope and decrease like the phase function of the Saturn dust. The slope of the function at intermediate scattering angles, or equivalently, the position of the SPF minimum, is connected to the reflectivity of dust grains and might be related to their composition and structure or to properties of their surfaces.

The HD 117214 SPF in Fig.~\ref{f_SPFs} shows the strongest similarity to those of the HD 35841 disk \citep{Esposito2018} and comet 67P/Churyumov–Gerasimenko \citep{Bertini2017}. The dust populations of these three objects might have a similar composition or effective size of particles. By fitting the HD 35841 data, \citet{Esposito2018} was unable to place strong constraints on either of them. The Mie models for the HD 35841 SPF preferred low-porosity grains (<12\%) that might be composed of carbon rather than astrosilicates and have roughly one-third water ice by mass. The authors considered these results to be of low significance, however. To fit the SPF of comet 67P, \cite{Moreno2018} used the T-matrix and geometric optics codes. The authors found that different types of oriented elongated particles with equivalent radii of 7 to 10 $\mu$m, porosity in the range of 60\% - 70\%, and refractive index of $m= 1.6 + 0.1i$ reproduce the coma dust SPF very well. 

As a final remark on Fig.~\ref{f_SPFs}, we note that for a proper comparison, the empirical SPFs should be normalized so that the integral of each phase function over 4$\pi$ steradians equals unity,
\begin{equation} \label{eq_SPF}
\int\limits_{0}^{2\pi} \int\limits_{0}^{\pi} SPF(\theta) \sin(\theta)\, d\theta\, d\phi = 1. 
\end{equation}
This normalization enables comparing the relative scattering efficiencies at each scattering angle between different dust populations. 

Another useful normalization is to plot the SPF as a ratio of dust surface brightness and stellar flux, expressing thus the phase function in terms of contrast or dust reflectance. Integration of the derived contrast curve (sometimes called ``albedo'' curve) over a full solid angle provides an estimate for the average scattering cross section and thus the average dust albedo. As mentioned above, the albedo is a characteristic from which we can retrieve additional information on the dust composition. 

However, an analysis like this requires an accurate measurement of the scattered flux in the entire range of scattering angles from 0 to $180^\circ$. With the exception of Solar System objects, no such measurement has been achieved because the smallest and largest angles are not observable in the debris disks surrounding other stars. This limits our knowledge of the dust reflectance in these systems. Nevertheless, the slope of the linear fit to the SPF measured at intermediate scattering angles, which indicates the position of the SPF minimum (below or above $\theta =90^\circ$), and the estimated contrast might shed light on the properties of dust grains in other stellar systems.

\subsection{Dynamical interactions between planets and the disk}
Similar to the HIP 79977 and many other debris systems, the HD 117214 disk shows a 0.4$''$ (43 au) cavity that most probably is cleared of dust. Since the discovery of extrasolar planet systems, planets have been considered as possible creators of wide gaps inside debris disks \citep[e.g.,][]{Wisdom1980, Faber2007, Shannon2016} because they dynamically interact with the planetesimals and dust and gravitationally attract or scatter the debris. In the following, we test this hypothesis for the inner clearing of the HD 117214 disk and consider a model of planetary architecture, where one or more planets reside in the gap between two planetesimal belts. \\
Based on the IR excess of the HD 117214 spectral energy distribution, the disk is best fit by a double-belt system \citep{Chen2014, Jang-Condell2015}, even if the contribution of the inner disk is quite marginal. Using the temperatures at which the IR excess peaks and a blackbody model for dust particles forming a thin ring, we retrieve a radius of $\sim$14 and $\sim$0.7 au for the warm and hot components, respectively. However, from the comparison with images of resolved disks in the literature, it emerged that the blackbody assumption underestimates the separation of the belts and a correction is needed. Thus, we multiplied both separations by the empirical $\Gamma$ factor \citep{Pawellek2015}, which accounts for the difference between the disk radius calculated with the equilibrium temperature assumption and the actual disk radius measured from the images. %  mainly depends on the luminosity of the central star: 
After this correction, the external belt is placed at $\sim45.5$~au, which agrees well with the position obtained from the resolved images of the disk, whereas the inner disk is placed at $\sim3$~au, behind the coronagraph.\\
In order to model the possible planetary systems responsible for the gap and, at the same time, remain compatible with our detection limits, we used the analytical method described in \cite{Lazzoni2018}. Planets on eccentric orbits are assumed to induce an eccentricity of the belt, and the models of the disk presented in this paper suggest that this is almost circular. We therefore considered only configurations with planets on circular orbits.\\

\begin{figure}
\centering
\includegraphics[height=7.5cm]{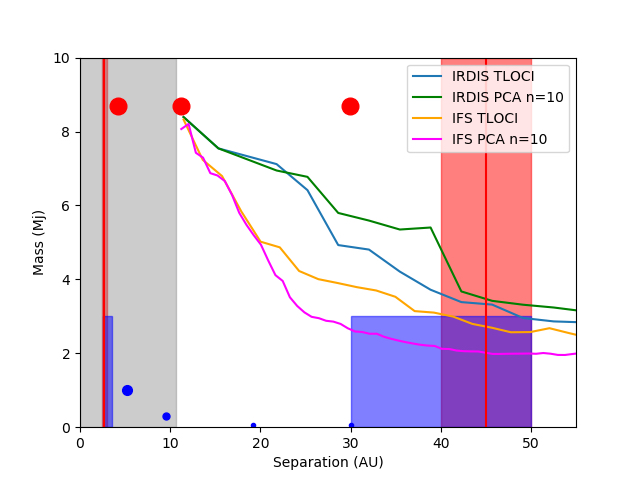}
   \caption{Detection limits for the companion mass in IRDIS and IFS datasets post-processed with TLOCI and PCA (10 modes). The red dots indicate the mass and semimajor axis of planets in the tested three planet configuration. The red vertical lines show the radial location of the inner and outer planetesimal belts, and the red shaded areas show their approximate extents. For comparison, the blue dots indicate the mass and semimajor axis of the Jovian planets, and the blue shaded areas show the extents of the main asteroid belt \citep[between 2.5 and 3.5 au;][]{Wyatt2008} and the Edgeworth–Kuiper belt \citep[between 30 and 50 au;][]{Stern1997} in the Solar System. The gray shaded region shows the IWA of the coronagraph. \label{f_contrastcurve}}
\end{figure}

Using the equation presented in \citet{Wisdom1980} for the region around the planet from which dust particles are scattered (chaotic zone), we can retrieve the mass and semimajor axis of one companion on a circular orbit that is responsible for carving the gap between 3 au and 40 au (corresponding to the positions of the inner and outer edges, respectively). However, when multi-planetary systems are considered, we also have to account for the stability of the configuration. We modeled a planetary system with two and three equal-mass planets on circular orbits, adding the hypothesis of maximum packing conditions so that the companions were as close as possible to preserve the stability of the configuration. Furthermore, we compared the results obtained from this analysis with the detection limits of SPHERE. In Figure~\ref{f_cc} we show the contrast curves at 5$\sigma$ level for the IRDIS and IFS instruments as obtained with the TLOCI and PCA (10 modes) post-processing techniques \citep{Zurlo2014, Mesa2015}. In Figure~\ref{f_contrastcurve} we convert the contrasts into planetary masses using the AMES-COND theoretical models \citep{Baraffe2003} and adopting an age of 10 Myr, assuming the system to be part of
the Lower Centaurus Crux.

\begin{figure*}
\centering
\includegraphics[width=17.5cm]{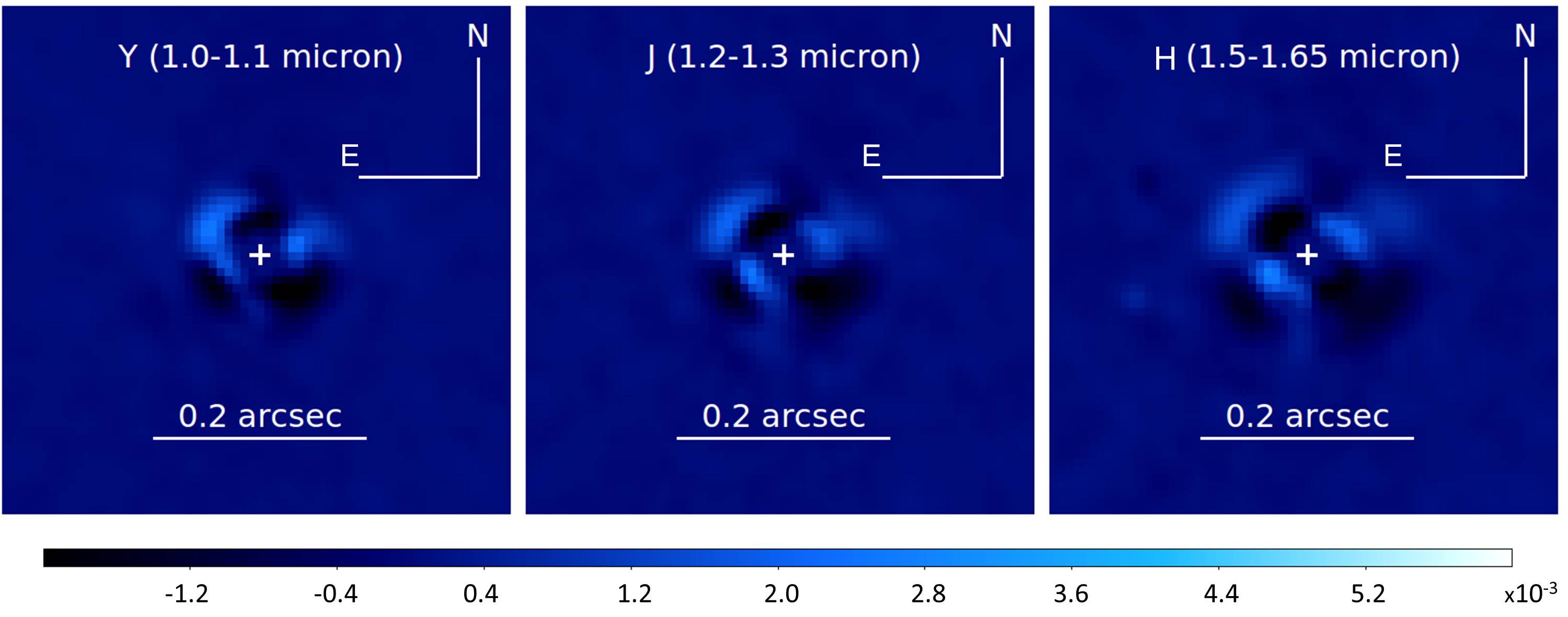}
   \caption{Final combination of the flux calibration images in the Y (\textit{left panel}), J (\textit{middle panel}), and H (\textit{right panel}) bands. The position of the star is marked by a white cross. The color bar shows counts per pixel. \label{f_final_H}}
\end{figure*}

\subsubsection*{Configuration with one or two companions}
For the assumptions of one or two equal-mass planets on circular orbits, we obtain companion masses higher than 80~$M_{\rm J}$. These values are in the stellar regime, which invalidates the equations we used. Massive companions like this would certainly have been detected if they were at projected radial separations larger than 0.11$''$. 

There is a possibility that a small stellar companion ($M > 0.08 M_\odot$) exists but has not been detected because it was hidden by the coronagraphic mask during the observation. This companion cannot be closer to the inner edge of the disk than the distance from the star given by the 2:1 resonance with the debris material. The modeled radius of the disk of 0.42$''$ (see Table~\ref{t_results}) implies that the semimajor axis of the orbit of this massive companion might be smaller than 0.26$''$. Assuming coplanarity of its orbit and a disk inclination of $\sim 70^\circ$, we conclude that the semiminor axis of the projected orbit of the companion might be smaller than 0.11$''$. This means that the companion may be behind the coronagraph during part of its orbit. 

In order to verify the presence of a stellar mass companion at projected separations smaller than 0.11$''$ in our data, we used the flux calibration frames recorded with the IFS at the start and end of the observation. The sky field rotation between these two exposures is about 31$^\circ$. We created two stellar flux images (image 1 and image 2) for each of the Y ($1.0 - 1.1$ $\mu$m), J ($1.2 - 1.3$ $\mu$m), and H ($1.5-1.65$ $\mu$m) bands from the IFS flux calibrations taken before and after the science sequence, respectively. Both images were normalized at their peak value, and their differential image was calculated. Figure~\ref{f_final_H} shows the final images of the Y, J, and H bands obtained as the difference of two differential images after their derotation: the first image is derotated by the parallactic angle of image 1, and the second image is derotated by the parallactic angle of image 2. After these steps, the static  aberrations are expected to cancel out in the final image (Fig.~\ref{f_final_H}), and a companion, if any exists, is expected to appear as a bright spot surrounded by two dark points. 

We did not detect a clear signal from a companion in the final images of the Y, J, and H bands. Therefore we computed the contrast reached at small radial separations ($< 0.14''$) in these images. For each of the radial separations listed in Col.~1 of Table~\ref{t_contrast} the standard deviation of the flux distribution was estimated in eight sectors of the concentric annulus of one pixel width. The limiting contrast (Cols.~$2-4$) was derived as the five-fold of the median standard deviation obtained for the corresponding annulus. The contrast values were converted into the planet mass using the AMES-COND models for the stellar ages of 10 (Cols.~8 and 9 in Table~\ref{t_contrast}) and 20 Myr (Cols.~10 and 11 in Table~\ref{t_contrast}). 

The same criterion for estimating the limiting magnitude was applied to the flux calibrations of the whole SHINE survey (660 datasets in total). This verification resulted in 28 detections of stellar companions, and only 6 of them were later rejected by a close examination. It is therefore a robust estimate of the limiting magnitude and allows us to place stringent constraints on possible very close companions to HD 117214. We can exclude, for instance, a stellar companion ($M > 0.08 M_\odot$) at a radial separation larger than 50 mas from the star for an age of 10 Myr, and from 75 mas for an age of 20 Myr. Adopting the older age, we may exclude a stellar companion at more than 16 au from the star, even if it were along the minor axis of the projected orbit at the time of the observation.

\subsubsection*{Configuration with three companions}
\begin{figure}
\centering
\includegraphics[width=7.5cm]{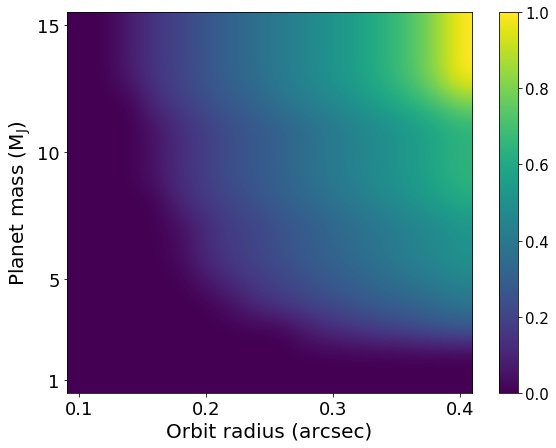}
   \caption{Detection probability map for the J-band data assuming an age of the system of 10 Myr. \label{f_dmapJ10}}
\end{figure}

For the configuration with three companions of equal mass, we obtained a more reasonable value of 8.7~$M_{\rm J}$ for the masses of planets on orbits with semimajor axes of $\sim$4.2, $\sim$11.2, and $\sim$29.9 au. We plot these values as red dots together with the contrast curves that we obtained for HD~117214 in Fig.~\ref{f_contrastcurve}. The innermost companion of this configuration is below the detection limits and would not have been detected in our observation. Although we are sensitive enough to the outermost planet, the detection probability of this companion is below 50\%. Because the system is inclined by 70$^\circ$, the planet might be at smaller projected separation at the time of the observation and therefore would remain undetected. In Fig.~\ref{f_dmapJ10} we show the probability detection map for 1 to 15 ~$M_{\rm J}$ planetary companions to HD 117214 orbiting a star on circular orbits with radii in the range between 0.1$''$ and 0.4$''$. The detection probability is defined as a fraction of the orbit with projected radial separations for which the companion contrast is above the contrast curve, and therefore it is model-dependent. The map shown in Fig.~\ref{f_dmapJ10} is derived for the stellar age of 10 Myr adopting AMES-COND models for the magnitudes of young giant planets in the J band.  

Given our sensitivity limits and detection probability, the three-planet configuration as well as a planetary system with more planets cannot be ruled out. 
In the last case, the planets might have significantly lower masses, as the comparison with the Jovian planets in Fig.~\ref{f_contrastcurve} shows. In the Solar System, four giant planets maintain a similar gap between the main asteroid belt and the Edgeworth–Kuiper belt. \\
Some caveats should be mentioned here. First of all, the disk may not have an inner hotter component. If this is the case, this analysis cannot be applied because a less massive and not detectable planet close to the belt may be sufficient to explain the shaping of the latter, and other mechanisms (such as Poynting-Robertson drag) may be efficient enough to clear the remaining inner part of the system from dust.\\
We also considered only equal-mass planets to avoid degeneracies in the solutions. We could then account for different configurations with, for example, a more massive inner companion that would be harder to detect due to its vicinity to the star and to the inclination of the system, and two smaller and undetectable planets placed farther away. 

\section{Summary} \label{s_Summary}
We presented the first images of scattered light (total and polarized intensities) from the debris disk around the Sco-Cen star HD 117214. The images were obtained at optical and near-IR wavelengths using the SPHERE instruments IRDIS, IFS, and ZIMPOL and have the highest resolution (up to 25 mas) achieved to date. 

The images reveal a bright axisymmetric debris ring with a radius of $\sim 0.42''$ ($\sim$~45~au at a distance of 107.6 pc) that is detected up to a radial distance of $\sim$1$''$ ($\sim$100~au). The disk is inclined at $71.0^{\circ} \pm 2.5^{\circ}$ and has a major axis at PA close to $180^{\circ}$ EoN. 

The west side of the disk is much brighter than the east side, in particular in the polarized intensity. This brightness asymmetry can be explained with the forward-scattering dust grains, implying that the west side is the near side of the disk. We modeled the observed brightness distribution with a 3D axisymmetric planetesimal ring and found that the underlying SPF can be best approximated with a linear combination of two HG functions: one describing the forward scattering with the asymmetry parameter $g=0.66,$ and the other describing backward scattering with $g=-0.22$. The results of the photometric analysis of the data are listed below.
\begin{itemize}
\item The total disk magnitude in polarized flux in the ZIMPOL VBB filter is 
$mp_{\mathrm{disk}}(\mathrm{VBB})$ = 16.48$^m \pm$ 0.30$^m$, whereas the stellar magnitude is 
$m(\mathrm{VBB})$ = 7.72$^m \pm 0.06^m$. This yields a disk-to-star contrast $(F_{\rm pol})_{\rm disk}/F_{\rm \ast}$ of $(3.1 \pm 1.2)\cdot 10^{-4}$.\\
The measured peak surface brightness of the polarized light is ${\rm SB} \mathrm{_{peak}(VBB)} = 15.1^m \pm 0.3^m$  arcsec$^{-2}$. This corresponds to a surface brightness contrast of 
${\rm SB} \mathrm{_{peak}(VBB)} - m \mathrm{_{star}(VBB)} = 7.4$ mag arcsec$^{-2}$. 
\item The comparison of the fractional polarized light flux $(F_{\rm pol})_{\rm disk}/F_{\rm \ast}$ in the VBB filter with the fractional IR luminosity of the disk $L_{\mathrm{IR}}/L_{\ast}$ yields $\Lambda = 0.12$. The same value was previously measured for an edge-on debris disk HIP 79977. The $\Lambda$ parameter together with the maximum polarization fraction $p_m$ provides an estimate for the ratio between the scattered-light luminosity and IR luminosity of the disk and thus for the disk-scattering albedo.
\item The comparison of the HD 117214 SPF measured in this work with the empirical SPFs of other debris disks and different dust populations in the Solar System shows that the forward-scattering peak of the SPF is well reproduced by Fraunhofer diffraction. Based on this comparison, we suggest that the average effective size of particles in the HD 117214 disk is about a few microns. 
\end{itemize}

We do not detect any planetary mass companion within the observed 40 au cavity, although we are sensitive down to $\sim$ 4 $M_{\rm J}$ planets at projected separations between 20 and 40 au in the IFS dataset. 
Outside of the planetesimal belt, we found 20 candidates in the IRDIS FoV, but they require a second epoch of observations in order to be confirmed and properly characterized.

\begin{acknowledgements}
We would like to thank the referee for many thoughtful comments that helped to improve this paper. 
Part of this work has been carried out within the framework of the National Centre for Competence in 
Research PlanetS supported by the Swiss National Science Foundation (SNSF). N.E. acknowledges the financial support of the SNSF. J.\,O. acknowledges financial support from the ICM (Iniciativa Cient\'ifica Milenio) via the N\'ucleo Milenio de Formaci\'on Planetaria grant, from the Universidad de Valpara\'iso, and from Fondecyt (grant 1180395). A.Z. acknowledges support from the CONICYT + PAI/ Convocatoria nacional subvenci\'on a la instalaci\'on en la academia, convocatoria 2017 + Folio PAI77170087. \\
This work has made use of the the SPHERE Data Centre, jointly operated by OSUG/IPAG (Grenoble), PYTHEAS/LAM/CESAM (Marseille), OCA/Lagrange (Nice), Observatoire de Paris/LESIA (Paris), and Observatoire de Lyon. 
SPHERE is an instrument designed and built by a consortium consisting
of IPAG (Grenoble, France), MPIA (Heidelberg, Germany), LAM (Marseille,
France), LESIA (Paris, France), Laboratoire Lagrange (Nice, France), 
INAF – Observatorio di Padova (Italy), Observatoire de Gen`eve 
(Switzerland), ETH Zurich (Switzerland), NOVA (Netherlands), ONERA (France) 
and ASTRON (Netherlands), in collaboration with ESO. 
SPHERE was funded by ESO, with additional contributions from CNRS (France), 
MPIA (Germany), INAF (Italy), FINES (Switzerland) and NOVA (Netherlands). 
SPHERE also received funding from the European Commission Sixth and 
Seventh Framework Programmes as part of the Optical Infrared 
Coordination Network for Astronomy (OPTICON)
under grant number RII3-Ct-2004-001566 for FP6 (2004–2008), grant number
226604 for FP7 (2009–2012) and grant number 312430 for FP7 (2013–2016).

\end{acknowledgements}

\bibliographystyle{aa} 

\bibliography{HD117214_reference.bib} 

\newpage 
\appendix
\section{HD~117214 polarimetric data per block of SP cycles} \label{s_QphiUphi_app}
\begin{figure*}
\centering
    \includegraphics[width=16cm]{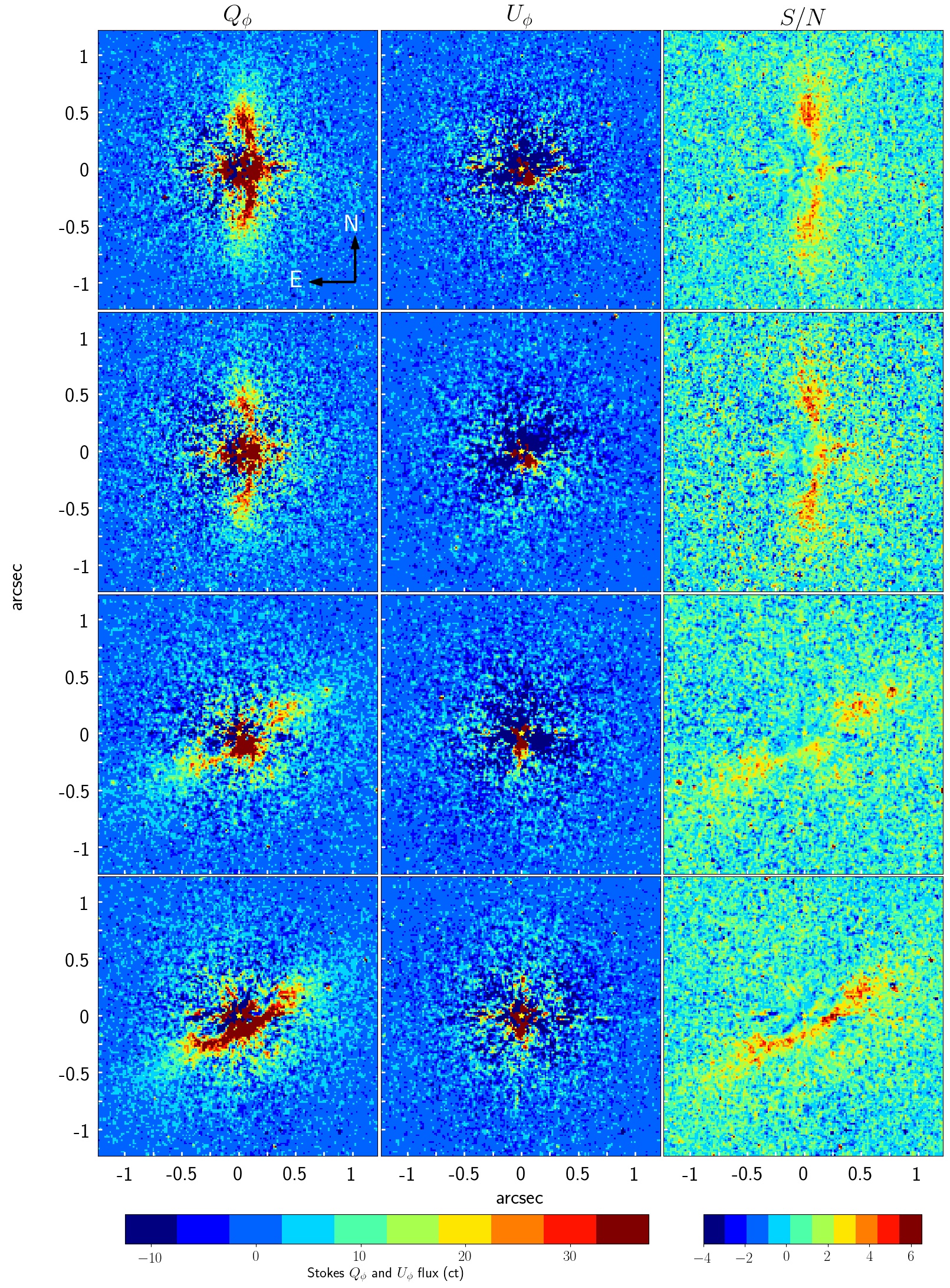} 
 \caption{$Q_\varphi$ (\textit{left column}) and $U_\varphi$ (\textit{middle column}) images and the S/N maps (\textit{right column}) of the polarimetric data per block of SP cycles. The first three rows show the data of first three blocks of SP cycles taken on 2018 February 28. Data recorded on 2018 June 22 are shown in the bottom row. Data in the lower two rows are taken with the sky field rotated by 60$^\circ$ on detector. \label{f_all_Qphi}}
\end{figure*}

Figure \ref{f_all_Qphi} shows the $Q_\varphi$ and $U_\varphi$ images and S/N map of different SP cycles recorded with the ZIMPOL. The highest quality data (top row) are taken at the beginning of the first observation on 2018 February 28 (see Table~\ref{t_Settings}) when the observing conditions were best (Sect.~\ref{s_Observations}). In these data, the scattered light from the disk is detected with the highest S/N, and the far side of the disk is visible on the northern extension. The images obtained from the data of the second and third SP blocks on February 28 are shown in the second and third rows. The lower quality of these images compared to the $Q_\varphi$ image in the top row is the result of degrading observing conditions, which led to the nondetection of the disk in the last SP cycles (fourth block) on that night. 

The images in the bottom row show the dataset from the second observing run on 2018 June 22. The $Q_\varphi$ and   $U_\varphi$ images presented in Fig.~\ref{f_QphiUphi} are the mean of the data shown in the top, second, and bottom rows of Fig.~\ref{f_all_app}. 
  
\section{Total intensity images (TLOCI, PCA) and detection limits on companions around HD~117214} \label{s_all_dr_app}
\begin{figure*}
\centering
    \includegraphics[width=17.5cm]{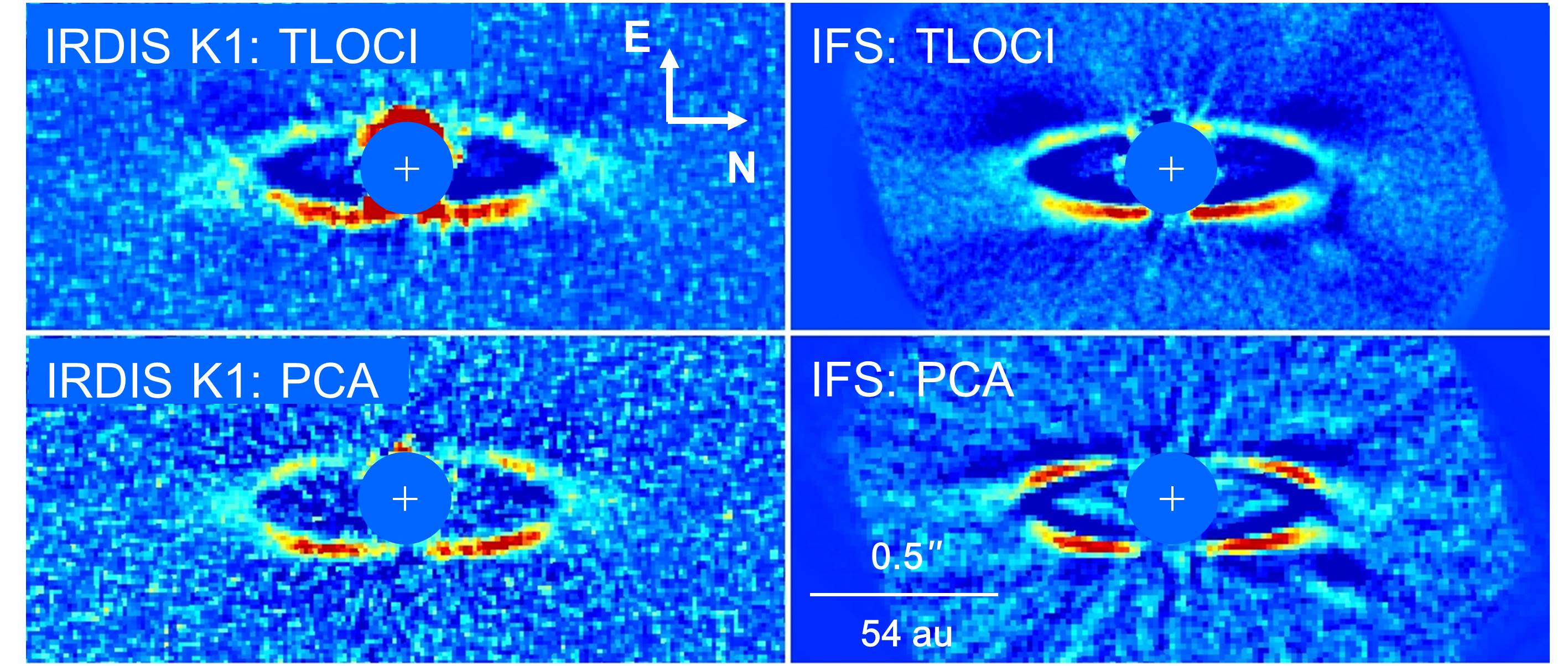} 
    \caption{Total intensity images of the HD 117214 debris disk obtained with TLOCI and PCA post-processing of the IRDIS K1 dataset (\textit{left column}) and spectrally combined IFS data (\textit{right column}). The PCA data reduction has 10 principal components. The position of the star is marked by a white cross. The color scale is in arbitrary units. \label{f_all_app}}
\end{figure*}

\begin{figure}
\centering
  \includegraphics[width=8.5cm]{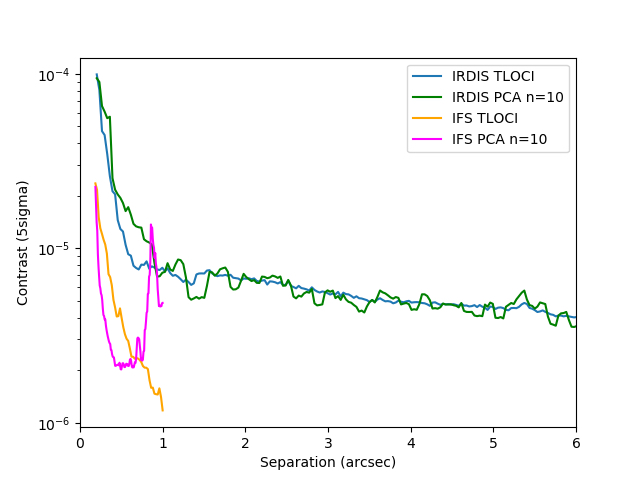} 
    \caption{Limits of detection on close companions around HD~117214 in 5$\sigma$ contrast. \label{f_cc}}
\end{figure} 

Figure \ref{f_all_app} shows the total intensity images obtained with IRDIS in K1 band (left column) and spectrally combined IFS data (right column) using different post-processing techniques. The 5$\sigma$ limits for the detection of companions in these images are displayed in Fig.~\ref{f_cc} as a function of angular separation. The contrast curves are corrected for the coronagraph throughput. For radial distances in the range $0.2'' < r < 0.8''$ , the contrast measured in the IFS dataset is significantly higher than in the IRDIS dataset due to the spectral diversity of the data. 

Table~\ref{t_contrast} contains the detection limits for stellar companions in the Y, J, and H bands reached at small radial separations ($< 0.14''$). The contrasts are converted into planet magnitudes (Cols.~$5-7$) and planet masses (Cols.~$8-10$) in J and H bands using the AMES-COND models for stellar ages of 10 and 20 Myr.

\begin{table*}  
 \caption[]{Limiting contrast and limits on the detectable companion mass from the flux-calibration data.  \label{t_contrast} }
          \centering
             \begin{tabular}{ccccccccccc}
            \hline
            \hline
            \noalign{\smallskip}
\multirow{3}{*}{Separation} & \multicolumn{3}{c}{Contrast} & \multicolumn{3}{c}{Companion detection limit} & \multicolumn{4}{c}{Companion mass limit}\\         
    &  &   & & & & & \multicolumn{2}{c}{10 Myr}&        \multicolumn{2}{c}{20 Myr}\\
     &  Y band   &  J band  & H band &  Y band &  J band  & H band &  J band  & H band & J band  & H band \\
(mas)  &  (mag) &   (mag)   &  (mag) &(mag)  &  (mag)   &        (mag) & ($M_{\rm J}$) & ($M_{\rm J}$) & ($M_{\rm J}$) & ($M_{\rm J}$) \\
            \hline
            \noalign{\smallskip}
37      & 5.24 & 5.24  &  4.74  &  7.44  &  7.25  &  6.55         &  126  & 141 & 180 & 201\\
45      & 5.62 & 5.83  &  5.47  &  7.82  &  7.84  &  7.28  &  78  & 87 & 122 & 126\\
52      & 5.80 & 6.04  &  6.03  &  8.00  &  8.05  &  7.84  &  71  & 53 & 104 & 88\\
60      & 6.17 & 6.08  &  6.36  &  8.37  &  8.09  &  8.17  &  69  & 44 & 102 & 71\\
67      & 6.67 & 6.22  &  6.35  &  8.87  &  8.23  &  8.16  &  61  & 44 & 94 & 71\\
75      & 7.24 & 6.61  &  6.28  &  9.44  &  8.62  &  8.09         &  45  & 46 & 72 & 75\\
82      & 7.88 & 7.33  &  6.40  &  10.08 &  9.34  &  8.21         &  22  & 43 & 44 & 69\\
90  & 8.30 & 7.83  &  6.59  &  10.50 &  9.84  &  8.40  &  18  & 30 & 23 & 62\\
97  & 8.55 & 8.21  &  7.03  &  10.75  &  10.22 & 8.84  &  16  & 22 & 18 & 45\\
104 & 8.64 & 8.24  &  7.52  &  10.84  &  10.25 & 9.33     &  16  & 19 & 18 & 29\\
112 & 8.63 & 8.37  &  8.07  &  10.83  &  10.38 & 9.88  &  16  & 16 & 17 & 17\\
119 & 8.56 & 8.36  &  8.39  &  10.76  &  10.37 & 10.20  &  16  & 15 & 17 & 16\\
127 & 8.63 & 8.45  &  8.52  &  10.83  &  10.46 & 10.33  &  16  & 14 & 16 & 15\\
134 & 8.67 & 8.58  &  8.49  &  10.87  &  10.59 & 10.30  &  15  & 14 & 16 & 15\\
142 & 8.95 & 8.88  &  8.43  &  11.15  &  10.89 & 10.24  &  14  & 14 & 15 & 15\\

  \noalign{\smallskip}
    \hline
 \end{tabular}\\
\end{table*}

\section{Posterior distributions of the fit parameters and residual images for the models with one and two HG parameters} \label{s_modeling_app}
To determine the best-fit parameter set for the total intensity image in the K1 band (Fig.~\ref{f_imaging}, left panel), we implemented the standard ensemble sampler using the MCMC technique, as proposed by \cite{Foreman-Mackey2013}. As discussed in Sect.~\ref{Modelling}, we ran the MCMC sampler twice using two different SPFs for our model: the HG function in the first sample, and a linear combination of two HG functions (Eq.~\ref{eq:phase func}) in the second sample. We used the uniform priors as specified in Col.~2 of Table~\ref{t_results} and investigated the parameter space with an ensemble of 1000 walkers, which perform a random walk with 4000 steps and fit the model to the data at each step. The noise map for the fit was calculated as the standard deviation of the flux distribution in concentric annuli of the total intensity image, excluding regions that contain disk flux.

\begin{figure*}
\centering
    \includegraphics[width=16cm]{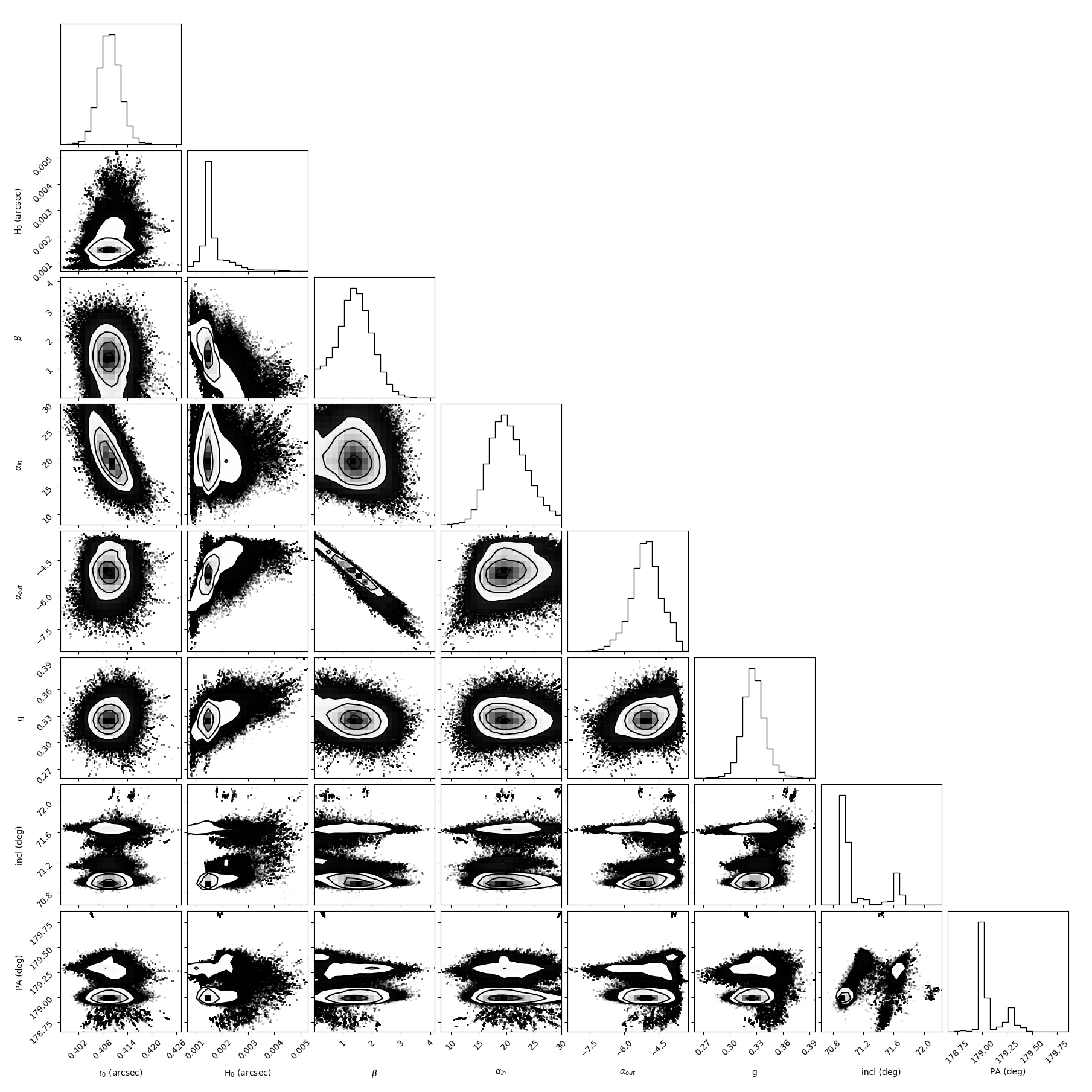} 

    \caption{Posterior distributions of the fit parameters for a model with one HG parameter. \label{f_mcmc_1g}}
\end{figure*}

\begin{figure*}
\centering
    \includegraphics[width=18cm]{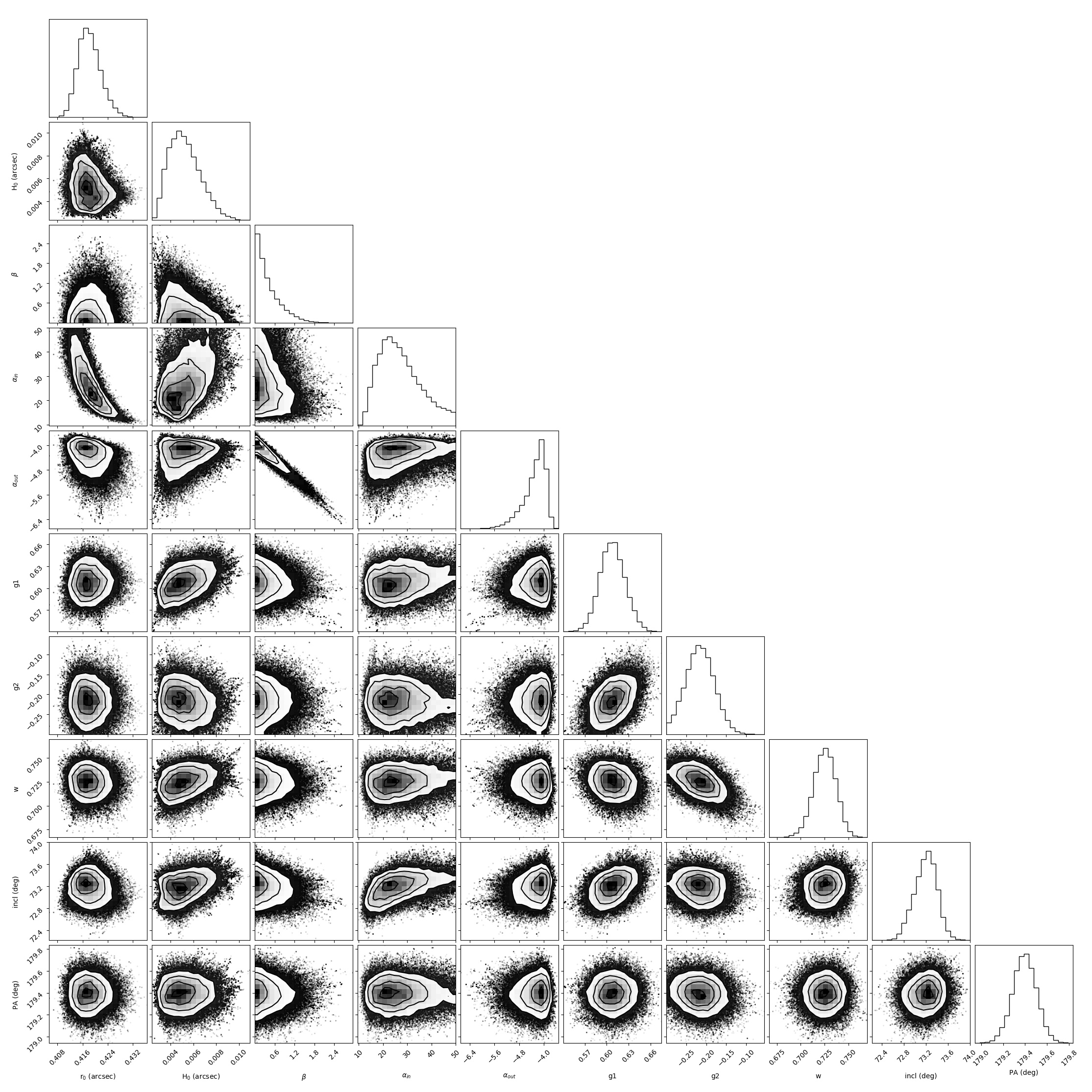} 
    \caption{Posterior distributions of the fitted parameters for a model with two HG parameters. \label{f_mcmc_2g}}
\end{figure*}

The maximum autocorrelation time of the parameter samples is 90 steps; therefore the first 400 steps were discarded to obtain the posterior distributions of parameters shown in Figs.~\ref{f_mcmc_1g} and~\ref{f_mcmc_2g}.  
The posterior distributions of PA and inclination of the disk for the model with one HG parameter (Fig.~\ref{f_mcmc_1g}) seem to have two local maxima. This bimodality does not disappear when the MCMC is run for a longer time.

\begin{figure*}
\centering
    \includegraphics[width=18cm]{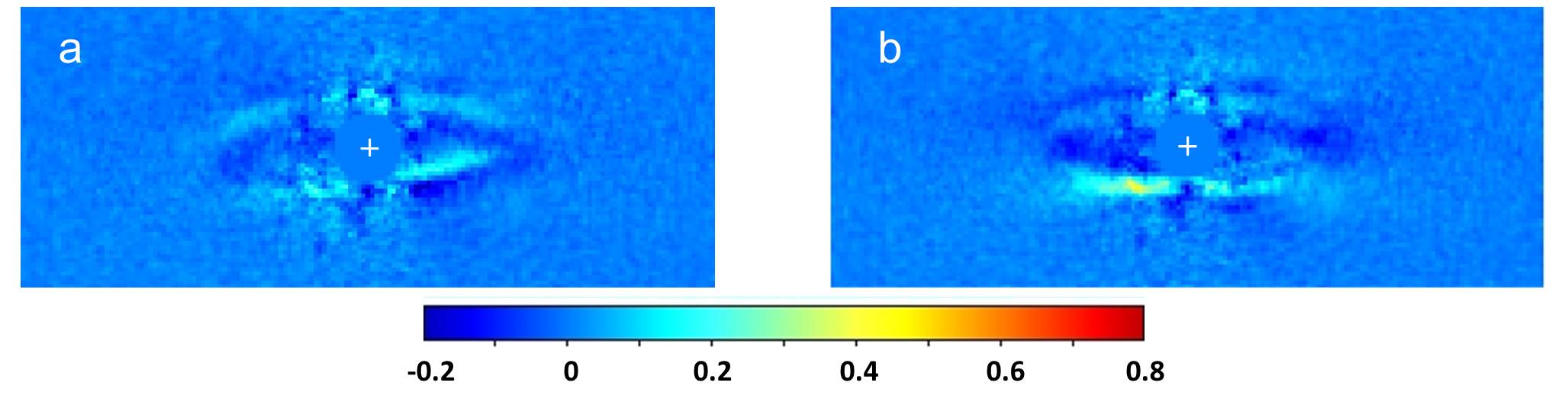} 
    \caption{Residual images of the models with two HG parameters (\textit{panel a}) and one HG parameter (\textit{panel b}). White crosses indicate the position of the star. The color bar shows the flux in counts per pixel. \label{f_residuals}}
\end{figure*}

With the best-fitting parameters found by each sampler (see Cols.~3 and 4 in Table~\ref{t_results}) we created best-fitting model images. Figure~\ref{f_residuals} shows the residual images of models with two HG parameters (panel a) and one HG parameter (panel b) obtained after subtracting the model images from the data (Fig.~\ref{f_model}a.)

\end{document}